\documentclass[twocolumn]{aastex631}

\newcommand{\surfbrightness}{421 }
\newcommand{\snfemb}{$-19.27 \pm 0.09$} 
\newcommand{\mugal}{$29.10 \pm 0.06$} 
\newcommand{\aN}{$25.110 \pm 0.064$}
\newcommand{\aM}{$22.809 \pm 0.053$}
\newcommand{\DSavg}{$12.981 \pm 0.082$} 
\newcommand{\DSMessier}{$13.001 \pm 0.127$} 
\newcommand{\hnaught}{$72.37 \pm 2.97$} 

\newcommand\plrnummiras{211}

\usepackage{amsmath}
\usepackage{multirow}
\usepackage{soul}

\begin{document}

\title{The Mira Distance to M101 and a 4\% Measurement of $H_0$}

\correspondingauthor{Caroline D. Huang}
\email{caroline.huang@cfa.harvard.edu}

\author[0000-0001-6169-8586]{Caroline D. Huang}
\affiliation{Center for Astrophysics $\vert$ Harvard \& Smithsonian, 60 Garden St., Cambridge, MA 02138, USA}

\author[0000-0001-9420-6525]{Wenlong Yuan}
\affiliation{Department of Physics and Astronomy, Johns Hopkins University, 3701 San Martin Dr., 
Baltimore, MD 21218, USA}

\author[0000-0002-6124-1196]{Adam G. Riess}
\affiliation{Department of Physics and Astronomy, Johns Hopkins University, 3701 San Martin Dr., 
Baltimore, MD 21218, USA}
\affiliation{Space Telescope Science Institute, 3700 San Martin Drive, Baltimore, MD 21218, USA}

\author{Warren Hack}
\affiliation{Space Telescope Science Institute, 3700 San Martin Drive, Baltimore, MD 21218, USA}

\author[0000-0002-4678-4432]{Patricia A. Whitelock}
\affiliation{South African Astronomical Observatory, P.O.Box 9, 7935 Observatory, South Africa.}
\affiliation{Department of Astronomy, University of Cape Town, 7701 Rondebosch, South Africa}

\author[0000-0001-6100-6869]{Nadia L. Zakamska}
\affiliation{Department of Physics and Astronomy, Johns Hopkins University, 3701 San Martin Dr., 
Baltimore, MD 21218, USA}

\author{Stefano Casertano}
\affiliation{Space Telescope Science Institute, 3700 San Martin Drive, Baltimore, MD 21218, USA}

\author[0000-0002-1775-4859]{Lucas M. Macri}
\affiliation{NSF’s NOIRLab, 950 N Cherry Ave, Tucson, AZ 85719, USA}

\author[0000-0001-9910-9230]{Massimo Marengo}
\affiliation{Department of Physics, Florida State University, 77 Chieftain Way, Tallahassee, FL 32306}

\author[0000-0003-4255-0767]{John W. Menzies}
\affiliation{South African Astronomical Observatory, P.O.Box 9, 7935 Observatory, South Africa.}

\author[0000-0003-4284-4167]{Randall K. Smith}
\affiliation{Center for Astrophysics $\vert$ Harvard \& Smithsonian, 60 Garden St., Cambridge, MA 02138, USA}


\begin{abstract}

The giant spiral galaxy M101 is host to the nearest recent Type Ia Supernova (SN 2011fe) and thus has been extensively monitored in the near-infrared to study the late-time lightcurve of the supernova. Leveraging this existing baseline of observations, we derive the first Mira-based distance to M101 by discovering and classifying a sample of \plrnummiras \ Miras with periods ranging from 240 to 400 days in the supernova field. Combined with new HST WFC3/IR channel observations, our dataset totals 11 epochs of \emph{F110W} (HST $YJ$) and 13 epochs of \emph{F160W}(HST $H$) data spanning $\sim$2900 days. We adopt absolute calibrations of the Mira Period-Luminosity Relation based on geometric distances to the Large Magellanic Cloud and the water megamaser host galaxy NGC 4258, and find $\mu_{\rm M101} = $ \mugal \ mag. This distance is in 1$\sigma$ agreement with most other recent Cepheid and Tip of the Red Giant Branch distance measurements to M101. Including the previous Mira-SNIa host, NGC 1559 and SN 2005df, we determine the fiducial SN Ia peak luminosity, $M^0_B =$ \snfemb \ mag. With the Hubble diagram of SNe Ia, we derive $H_0 =$ \hnaught \ km s$^{-1}$Mpc$^{-1}$, a $4.1\%$ measurement of $H_0$ using Miras. We find excellent agreement with recent Cepheid distance ladder measurements of $H_0$ and confirm previous indications that the local universe value of $H_0$ is higher than the early-universe value at $\sim$ $95\%$ confidence. Currently, the Mira-based $H_0$ measurement is still dominated by the statistical uncertainty in the SN Ia peak magnitude.

\end{abstract}

\section{Introduction} \label{sec:intro}

\begin{figure*}
\centering
\includegraphics[width=1.0\textwidth]{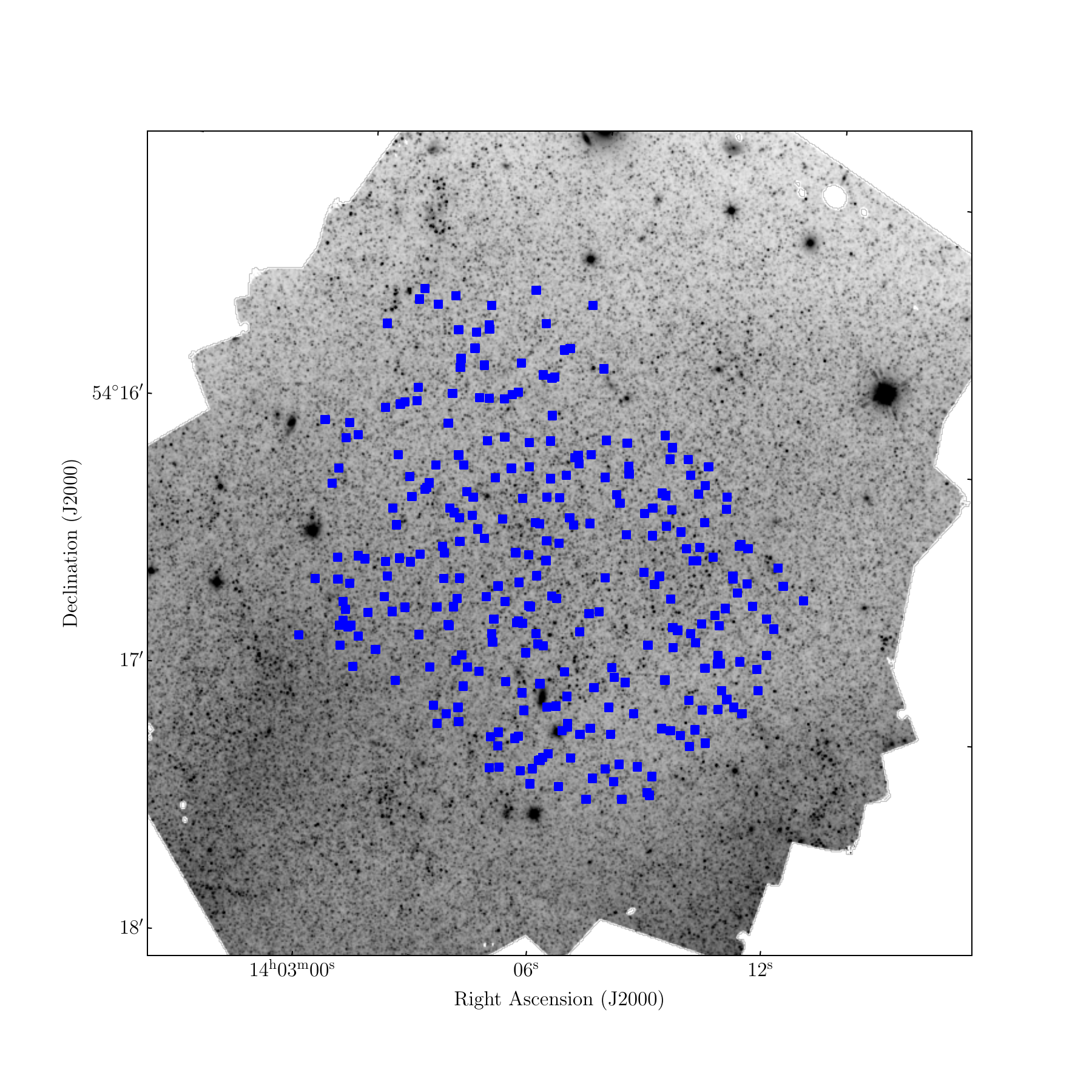}
\caption{Locations of Miras (blue squares) discovered using the selection criteria outlined in \S\ref{sec:selection_criteria}, plotted on the \emph{F160W} stacked image comprising 47 stacked individual exposures. 
\label{fig:mira_locations}}
\end{figure*}

Distances play a fundamental role in astronomy by allowing us to convert observed fluxes and angular separations into physical luminosities and sizes. Through the use of distance ladders, we can then extend these nearby geometric calibrations to cosmologically-relevant scales \citep{Riess_1998, Perlmutter_1998}. 

As one of the only cosmological parameters that can be measured in a model-independent manner, the rate of expansion of the universe at the present day, the Hubble constant ($H_0$), offers a unique opportunity to test the physical underpinnings of the standard Lambda Cold Dark Matter ($\Lambda$CDM) model of cosmology. Recent comparisons of the direct local determinations of $H_0$ (``Late" Universe), and the values predicted using cosmological model calibrated observations of the Cosmic Microwave Background (CMB; ``Early" Universe), show a persistent $\sim$9\% difference, which has become known as the ``Hubble tension.'' The most precise local measurement \citep[SH0ES,][]{Riess22} disagrees with the most precise value inferred from the early Universe \citep{Planck18VI} by $>$5$\sigma$, which may be an intriguing indication of new physics \citep[see reviews by][]{Di_Valentino_2021, Perivolaropoulos_Skara_2022, Kamionkowski_and_Riess, Hu_and_Wang_2023}. 

\emph{Hubble} Space Telescope (HST) observations of the Leavitt Law of Cepheid variables have been used with great success to calibrate the distances to nearby Type Ia supernovae (SNe Ia) in distance-ladder determinations of $H_0$ \citep[e.g.,][]{Freedman01, Riess22}. The uncertainty on the most precise Cepheid-based measurement \citep{Riess22} is now approaching 1\%, but the tension has only grown as the measurement uncertainties have decreased. Thus, independent calibrations of SNe Ia can provide a cross-check of the systematics and strengthen the evidence for additions to the Standard Model of Cosmology ($\Lambda$CDM). To this end, a number of other intermediate distances indicators have been used as alternatives to calibrate SNe Ia, including the Tip of the Red Giant Branch (TRGB) and Oxygen-rich (O-rich) Mira variables. 

At present, TRGB-based measurements of $H_0$ made with several different methods \citep{Freedman19, Anand22, Anderson_2023, Scolnic23} have slightly larger uncertainties and range of values than Cepheid-based measurements. The results of \cite{Anand22}, \cite{Anderson_2023}, and \cite{Scolnic23} agree with each other and the SH0ES measurement to within 1$\sigma$. The \cite{Freedman19} TRGB result from the Carnegie-Chicago Hubble Program (CCHP) prefers a somewhat lower value and sits between the Early and Late Universe measurements. An analysis by \citet[][see their Table 5]{Scolnic23} traces most of the differences in $H_0$ in these TRGB-based studies to differences in their treatment of SNe~Ia, totaling 2.0 km s$^{-1}$ Mpc$^{-1}$.

In \cite{Huang_2020}, we discovered and classified Mira variables --- fundamentally-pulsating asymptotic giant branch stars --- in NGC 1559, host to SN 2005df. We then used this sample to derive a calibration of the SN~Ia fiducial luminosity and a measurement of $H_0$ using the single SN~Ia calibrator. This distance ladder was composed solely of geometry, Miras, and SN~Ia, and thus, was completely independent of Cepheids or TRGB. This resulted in a measurement of H$_0 = 73.3 \pm 4.0$ km s$^{-1}$ Mpc$^{-1}$, which, due to its large uncertainties, is within 1$\sigma$ of $H_0$ derived using Cepheids and 1.5$\sigma$ of the value of $H_0$ inferred by \cite{Planck18VI}. The $\sim$ 5.5\% uncertainty of this measurement is dominated by the statistical uncertainty in the peak magnitude of the sole SN~Ia calibrator. 

Significantly improving the precision of this measurement requires calibrating additional SNe Ia. While well-sampled light curves generally reduce the uncertainty in the peak luminosity of an SN~Ia, there is typically still a $\sim 0.10$ mag uncertainty in the peak magnitude after standardization. Even with the inclusion of spectroscopy, \cite{Murakami_2023} found an overall $\sim0.12$ mag level of scatter remained for SNe Ia in the distance ladder. Thus, in this work, we focus on reducing the uncertainties on the Mira-based $H_0$ by calibrating a second SN~Ia, SN 2011fe, located in the nearby giant spiral galaxy M101.

Asymptotic giant branch (AGB) stars are a natural choice for standard candles since they are typically the brightest stars in old and intermediate-age populations.  Their luminosities allow them to bridge the gap between geometric measurements and the local host galaxies of SNe Ia, which are typically within $40$ Mpc. While AGB pulsation can be classified into several modes, and many types of AGB stars have long periods ($P > 30$ days) and follow Period-Luminosity Relations (PLRs), fundamentally-pulsating Mira stars are arguably the easiest subset of long-period variables to classify thanks to their large amplitudes of pulsation. They have long but stable periods which can be recovered with multi-epoch data covering several phase points and spanning at least one cycle of pulsation. Their pulsational periods and luminosities have been known to be related for several decades \citep{GlassLE81}. Amongst Miras, the short period ($P < 400$ days), Oxygen-rich (O-rich) subgroup are the most commonly employed for the cosmic distance ladder due to their small intrinsic scatter at near-infrared and longer wavelengths \citep{Yuan17a, Yuan17b, Huang_2020}. The more massive Carbon-rich (C-rich) Miras are also known to follow PLRs, particularly in the mid-infrared \citep[e.g.,][]{Iwanek_2021}. C-rich AGB stars have also been proposed collectively as a separate indicator \citep[JAGB;][]{Madore_2020}. 

\begin{figure*}
\includegraphics[width=\textwidth]{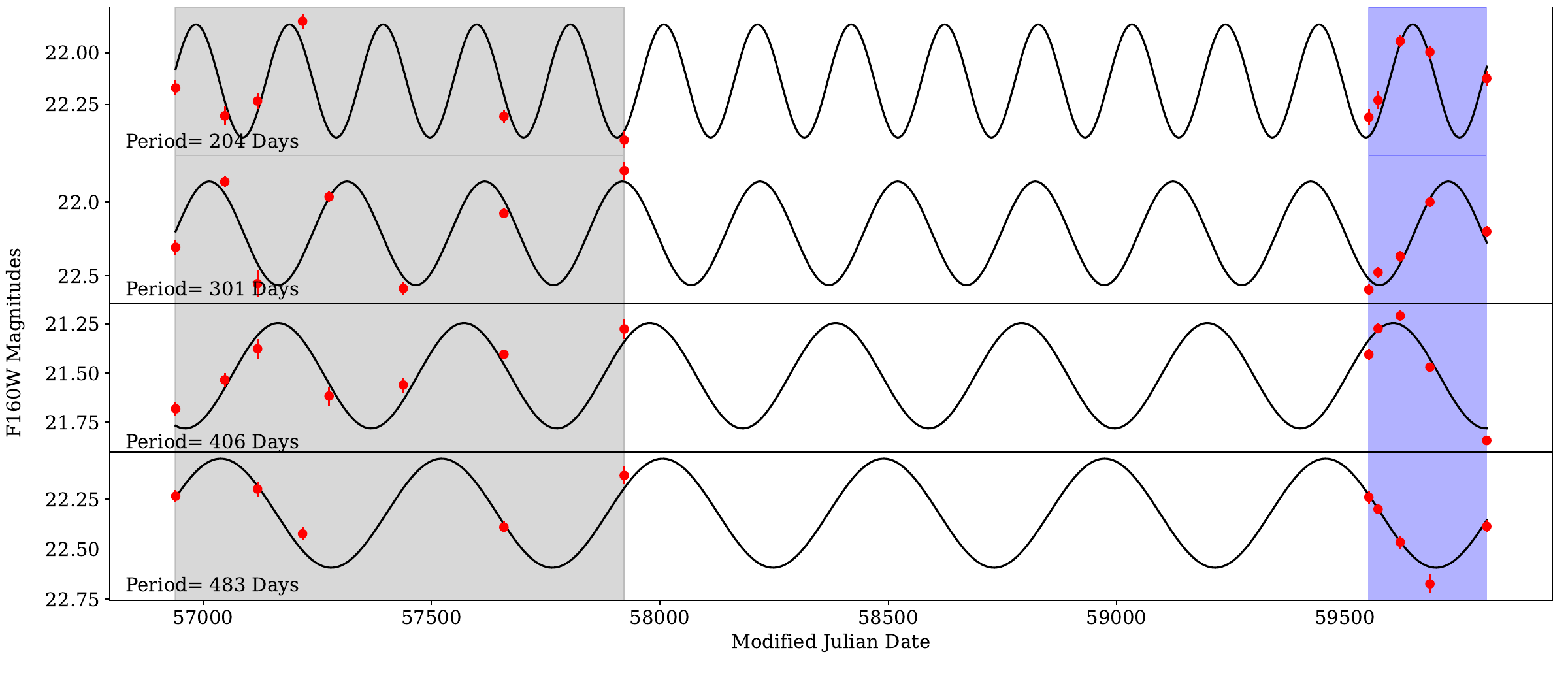}
\caption{Four representative light curves of candidate Miras with periods ranging from 204-472 days. The sinusoidal fit is shown in black and observations in red. Gray-shaded regions indicate the span of time of the archival observations while blue shaded regions span the new Cycle 29 observations. 
\label{fig:four_lcs}}
\end{figure*}


Mira variables are highly-evolved, fundamental-mode, thermally-pulsing stars at the tip of the asymptotic giant branch (AGB). In the near-infrared and longer wavelengths, they generally follow a tight ($\sigma \sim 0.13$ mag in $H$) PLR similar to the Leavitt Law for Cepheids. They have large amplitudes (frequently $\Delta V > 2.5$ mag or $\Delta K > 0.4$ mag are used) and long periods generally ranging from $\sim 100 - 2000$ days \citep[e.g.][]{Whitelock06, Soszynski09a, Samus_2017}. Miras can be classified as O-rich or C-rich depending on their surface chemistries. Stars entering the AGB will begin as Oxygen-rich; during this evolutionary phase, carbon synthesized through triple-alpha reactions is dredged up to the surface of the star, gradually increasing the photospheric C/O ratio. If this ratio exceeds 1, the star is classified as Carbon-rich. The relative abundances of carbon and oxygen on the surface of a star will determine the molecular features present in its spectra. Thus, C-rich and O-rich Miras do not always follow the same PLR, and separation of the two spectral subtypes is important to reduce scatter and bias of the PLR. 

The Miras used in the cosmic distance ladder are short-period ($P < 400$ d) O-rich variables since this subgroup has simpler physics, is the easiest to isolate, and follows the tightest PLR at the near-infrared wavelengths accessible to HST. The majority of short-period Miras are O-rich, which limits C-rich Mira contamination in the PLR.  At longer periods, some O-rich Miras may have increased luminosity due to hot-bottom burning (carbon-burning at the base of the outer convective layer) and will lie above the PLR \citep{Whitelock03}. Restricting the period range to shorter periods also has the benefit of enabling a shorter baseline of observations for characterization than would be necessary to determine the periods of the more slowly-varying Miras.

\begin{deluxetable*}{lllrccc}
\tabletypesize{\scriptsize}
\tablecaption{Summary of \emph{HST} Observations}
\tablewidth{0pt}
\tablehead{\colhead{Epoch} & \colhead{Date} & \colhead{Filter} & \colhead{Exposure (s)} & \colhead{No. Dithers} & \colhead{Proposal ID} & \colhead{Dataset}
}
\startdata
    1 & 2014-10-09 & \emph{F160W}& 1398.46 & 2 & 13737 & ICOY01 \\
    1 & 2014-10-09 & \emph{F110W} & 1598.47 & 2 & 13737 & ICOY01 \\
    2 & 2015-01-24 & \emph{F160W}& 827.72 & 3 & 13824 & ICL0A1 \\
    3 & 2015-04-06 & \emph{F160W}& 1398.46 & 2 & 13737 & ICOY02 \\
    3 & 2015-04-06 & \emph{F110W} & 1598.47 & 2 & 13737 &  ICOY02 \\
    4 & 2015-07-14 & \emph{F160W}& 1398.47 & 2 & 13737 & ICOY04 \\
    4 & 2015-07-14 & \emph{F110W} & 1598.47 & 2 & 13737 & ICOY04 \\
    5 & 2015-09-10 & \emph{F160W}& 1655.45 & 6 & 13824 & ICL0A2 \\
    6 & 2016-02-19 & \emph{F160W}& 1398.46 & 2 & 14166 & ICZJ01 \\
    6 & 2016-02-19 & \emph{F110W} & 1598.47 & 2 & 14166 & ICZJ01 \\
    7 & 2016-09-27 & \emph{F160W}& 8795.40 & 6 & 14166 & ICZJ03 \\
    7 & 2016-09-27 & \emph{F110W} & 5195.40 & 6 & 14166 & ICZJ03 \\
    8 & 2017-06-17 & \emph{F160W}& 1911.74 & 4 & 14678 & ID6V02 \\
    8 & 2017-06-17 & \emph{F110W} & 4011.74 & 4 & 14678 & ID6V02\\
    9 & 2021-12-03 & \emph{F160W}& 1311.76 & 4 & 16744 & IEQ601 \\
    9 & 2021-12-03 & \emph{F110W} & 1211.74 & 4 & 16744 & IEQ601 \\
    10 & 2021-12-24 & \emph{F160W}& 1311.76 & 4 & 16744 & IEQ602 \\
    10 & 2021-12-24 & \emph{F110W} & 1211.74 & 4 & 16744 & IEQ602 \\
    11 & 2022-02-10 & \emph{F160W}& 1311.76 & 4 & 16744 & IEQ603 \\
    11 & 2022-02-10 & \emph{F110W} & 1211.74 & 4 & 16744 & IEQ603 \\
    12 & 2022-04-16 & \emph{F160W}& 1311.76 & 4 & 16744 & IEQ604 \\
    12 & 2022-04-16 & \emph{F110W} & 1211.74 & 4 & 16744 & IEQ604 \\
    13 & 2022-08-19 & \emph{F160W}& 1311.76 & 4 & 16744 & IEQ605 \\
    13 & 2022-08-19 & \emph{F110W} & 1211.74 & 4 & 16744 & IEQ605 
\enddata
\tablecomments{All \textit{HST} observations analyzed in this paper. Full data set can be accessed at:\dataset[10.17909/z80p-1q97]{https://doi.org/10.17909/z80p-1q97}}
\label{tab:observations}
\end{deluxetable*}
The rest of the paper will be structured as follows. In \S\ref{sec:obsredpho} we discuss the observations, data reduction, and photometry process. In \S\ref{sec:Methods} we explain the variable search process and the selection criteria applied to obtain the Mira sample.  In \S\ref{sec:systematics} we discuss the potential impact of extinction and metallicity on our results. In \S\ref{sec:H0} we outline the individual rungs of the distance ladder, the derivation of $H_0$, and the error budget. In \S\ref{sec:results} we present and discuss the results and compare our measurement of the distance to M101 with previous distances obtained using Cepheids and TRGB. Finally, we conclude in \S\ref{sec:conclusions}. Throughout the paper, we use ``amplitude" to refer to the total (peak-to-trough) variation of a star's lightcurve as determined from the coefficients of the sinusoidal fit. 

\section{Observations, Data Reduction, and Photometry} \label{sec:obsredpho}
\setlength{\tabcolsep}{1em}

\begin{figure}
\includegraphics[width=0.5\textwidth]{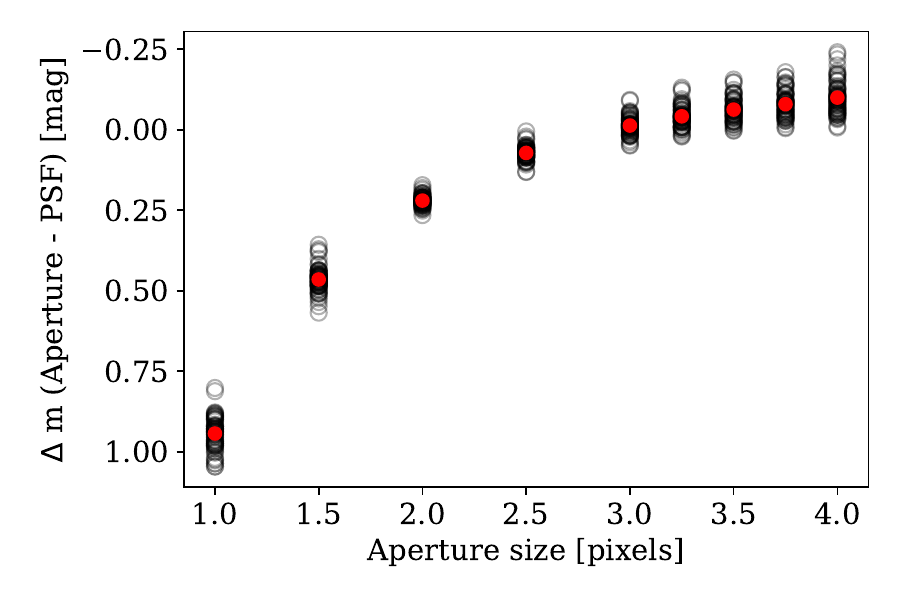}
\caption{The difference in aperture and PSF magnitudes for $\sim 70$ bright, isolated stars in our field. The difference in magnitude for the individual stars are given by the black open circles, while the final aperture correction derived from these measurements is shown in red. Uncertainties for the final aperture corrections are smaller than the individual points. 
\label{fig:growth_curve}}
\end{figure}

\subsection{Observations}\label{sec:obs}
Due to the long period of variability of Miras (around $100-400$ days for our intended targets), we found it advantageous to use a combination of archival and new observations to create the longest possible baseline. As the host to the nearest modern Type Ia Supernova, SN 2011fe, the supernova field of M101 was observed eight times over the course of three years in the near-infrared to study the late-time light curve of the supernova. Thus, we chose to target this region of the galaxy for additional Mira observations which improved the sampling and increased the area of the observations. With a combination of archival and new, more recent time-series, we are able to both increase the observational footprint and extend the observational baseline, doubling the number of recovered variables over using just the archival observations. The field that we targeted and the locations of the Miras discovered in the galaxy using the selection criteria from \S\ref{sec:Methods} are shown in Figure \ref{fig:mira_locations}. A summary of these observations and their exposure times can be found in Table \ref{tab:observations}. 


We use archival observations of the galaxy acquired in four separate proposals (GO-13737, PI: Shappee; GO-13824, PI: Kerzendorf; GO-14166, PI: Shappee; and GO-14678, PI: Shappee) in 2014-2017 using the Infrared (IR) channel of the Wide Field Camera 3 (WFC3) instrument on the \emph{Hubble} Space Telescope (HST) in two filters: \emph{F110W} and \emph{F160W}(the HST wide $YJ$ and wide $H$ filters, respectively). All of the observations were centered on or near the location of SN 2011fe ($\alpha= 14^{\rm h}03^{\rm m}05.711^{\rm s}$, $\delta = +54^\circ16'25.22''$, Equinox:J2000). The number of dithers ranged from 2 to 6 and the cadence for these observations is uneven, ranging from 58 days to 220 days with a total baseline of 982 days. These data have not previously been analyzed in order to search for Miras or other periodic variables. 

\begin{figure}
\includegraphics[width=0.5\textwidth]{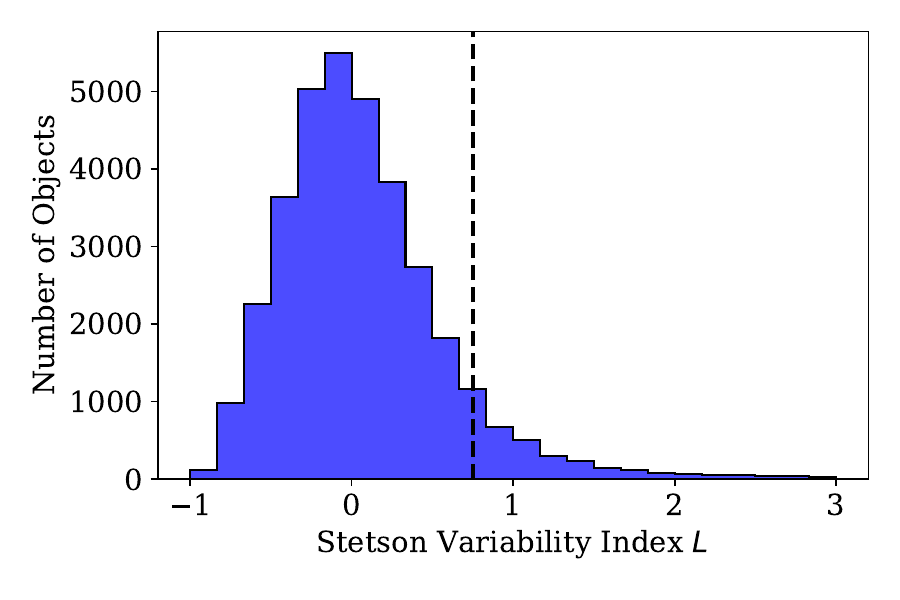}
\caption{Histogram of Welch-Stetson variability index $L$ values for all sources with at least 10 epochs of observation -- a total of 35,000 objects. The dashed black line indicates the $L$=0.75 threshold we used to begin our variability search. 
\label{fig:Lhist}}
\end{figure}

\begin{figure*}
\includegraphics[width = 0.33\textwidth]{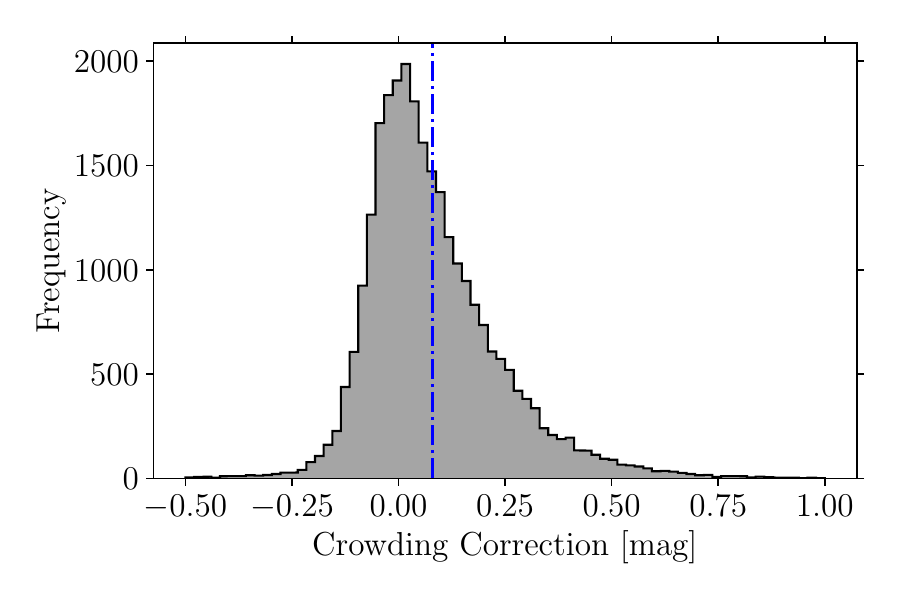}
\includegraphics[width = 0.33\textwidth]{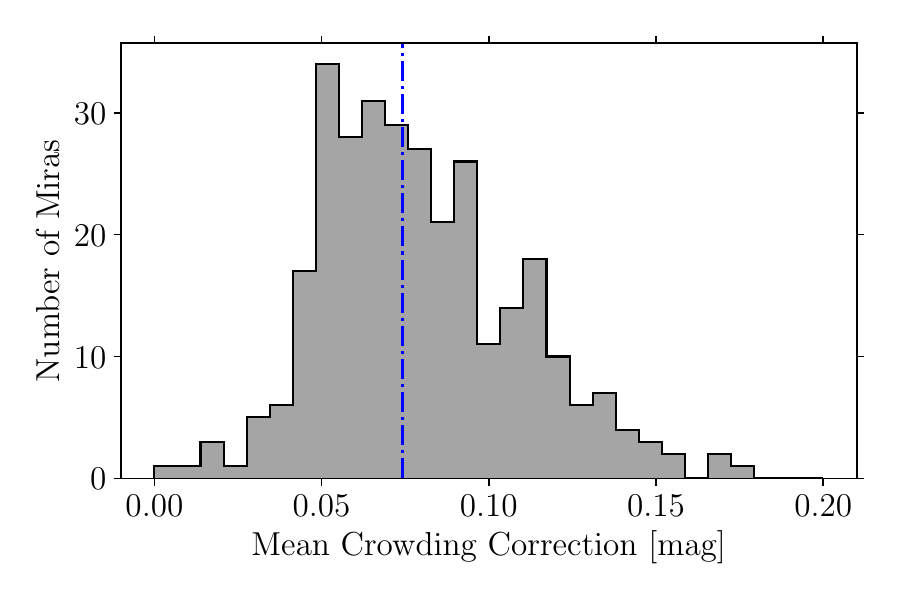}
\includegraphics[width=0.33\textwidth]{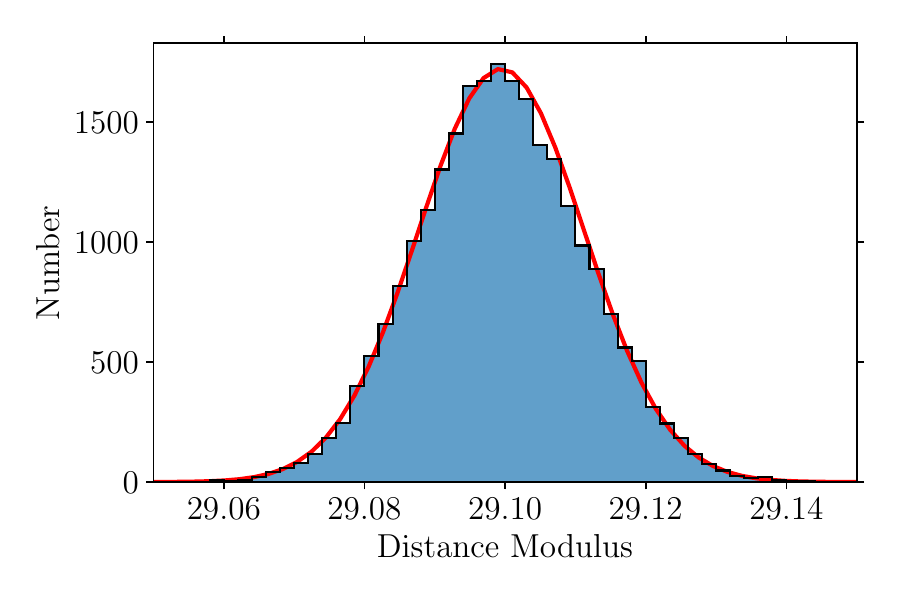}
\caption{{\bf Left:} Distribution of all crowding corrections from every run of the artificial star tests. The peak of this distribution is close to zero and depends on the binning chosen. {\bf Middle:} Distribution of the mean crowding correction for each Mira. The blue dashed line in both plots indicates the mean of the distribution, $0.080$ mag and $0.074$ mag, respectively. {\bf Right:} Distribution of zeropoints for the 260 days $< $ P $<$ 400 day sample obtained through bootstrap Monte Carlo application of crowding corrections. Histogram of the Monte Carlo in blue, Gaussian distribution in red. 
\label{fig:crowding}}
\end{figure*}

\begin{figure}[ht!]
\plotone{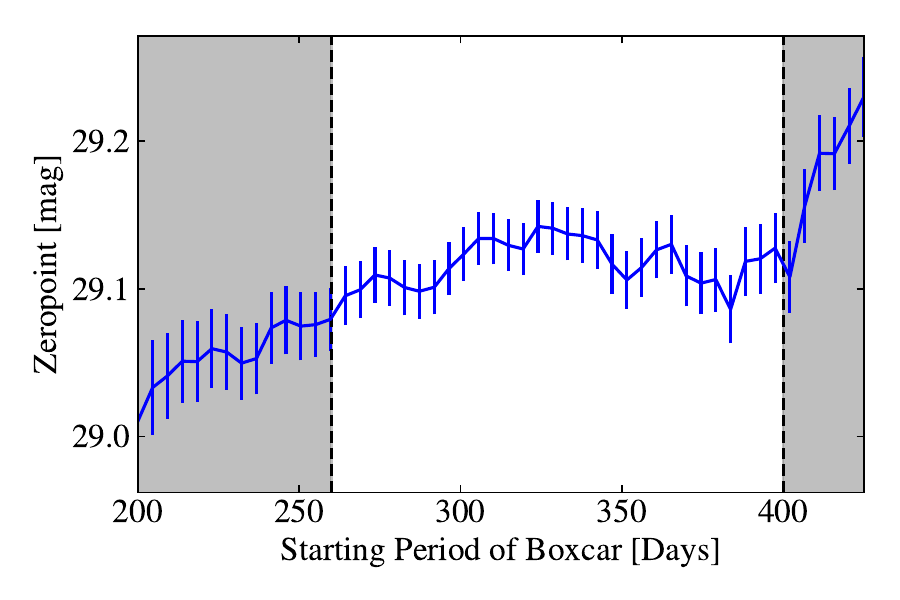}
\caption{Zeropoint as a function of period using a boxcar fit with a width of 75 days. The gray shaded regions show the period ranges that are excluded in our best-fit analysis. 
\label{fig:boxcar}}
\end{figure}


In addition, we acquired five new epochs of HST WFC3/IR \emph{F110W} and \emph{F160W} imaging (GO-16744, PI: Huang) from 2021 December to 2022 August.  For each filter and epoch there are  four sub-pixel-dithered exposures. These observations are centered on the supernova and span a baseline of $258$ days with an uneven, approximately power-law cadence varying from 20 days to 124 days that was designed to reduce aliasing \citep{Freedman_1994}. The more recent observations were timed to improve recovery of the short period Miras, allowing us to observe at least 3 phase points in one cycle for all periods $> 100$ days. Combined with the archival observations, the total baseline is 2871 days. Regions close to the supernova, with maximum overlap from the individual exposures, have up to 13 epochs of observations. Figure \ref{fig:four_lcs} shows the full baseline of archival and recent observations for selected variables with periods ranging from $200-500$ days.

\subsection{Filter Selection}

Previous observations of Miras in NGC 1559 \citep{Huang_2020} and NGC 4258 \citep{Huang18} also used \emph{F160W}for the Mira P-L relation. However, due to its close match in wavelength with the ground-based $J$-band filter, GO-13445 (PI: Bloom) used  \emph{F125W} instead of \emph{F110W} to obtain $J$-band observations in NGC 4258. In \cite{Huang18}, we discovered that the zeropoint of the Mira P-L relation remained the same regardless of whether a cut based on \emph{F125W}-\emph{F160W}color or \emph{F814W} variability was applied during the selection process for NGC 4258.  This issue is discussed in greater detail in \S\ref{sec:crich}, but the near-infrared wide filters encompass molecular features in the spectra of both subtypes, thus limiting our ability to separate O- and C-rich Miras using only near-infrared colors derived from HST wide bands. The \emph{F814W} filter should improve this separation, but even using \emph{F814W} variability as a selection criterion in \cite{Huang18} did not change the zeropoint by more than $\sim 0.02$ mag compared to using only F160W, which may suggest low C-rich contamination, or similarity of magnitudes for C-rich and O-rich Miras in HST NIR bands. Thus, only \emph{F160W} observations were obtained for NGC 1559. 

In this work, we attempt once again to separate C- and O-rich Miras on the basis of their colors in addition to the O- and C-rich separation and other criteria developed in \cite{Huang18, Huang_2020} which are discussed in greater detail in \S\ref{sec:selection_criteria}. We have chosen to use \emph{F110W} because of its greater wavelength separation from the \emph{F160W} filter and wider wavelength coverage (to improve signal-to-noise) compared to the \emph{F125W} filter used in the previous Mira study. The \emph{F110W} filter extends from 9000 to 14000~\AA, overlapping with the \emph{F814W} by about 600 \AA, while the \emph{F125W} filter extends from 11000 to 14000 \AA. This allows us to achieve increased signal-to-noise and somewhat better C- and O-rich separation using \emph{F110W}. 

\subsection{Data Reduction and Photometry}\label{sec:dataredux}

\setlength{\tabcolsep}{1em}
\begin{deluxetable*}{lccc}
\tabletypesize{\scriptsize}
\tablecaption{Mira Sample Selection Criteria}
\tablewidth{0pt}
\tablehead{ \colhead{} & \colhead{M101} & \colhead{NGC 1559} & \colhead{NGC 4258 (gold)}}
\startdata
   Period Cut: &  200 d $< P <$ 500 d  & $240$ d $< P < 400$ d &  $P < 300$ d \\
    Amplitude Cut: & $0.4 \text{ mag} < \Delta \emph{F160W} < 0.8$ mag & $0.4 \text{ mag} < \Delta \emph{F160W} < 0.8$ mag & $0.4 \text{ mag} < \Delta \emph{F160W} < 0.8$ mag \\
    Color Cut: & $m_{\emph{F110W}} - m_{F160W} < 1.3$ mag & --- &$m_{F125W} - m_{F160W} < 1.3$ mag  \\
    Surface Brightness Cut: & ---  & \surfbrightness counts/second &  --- \\
    $F$-statistic: & $\chi^2_{\text{s}}/\chi^2_{\text{l}} < 0.5$ & $\chi^2_{\text{s}}/\chi^2_{\text{l}} < 0.5$ & --- \\
    \emph{F814W} Detection: & ---  & --- & Slope-fit to \emph{F814W} data$ > 3\sigma$  \\
    \emph{F814W} Amplitude: &--- &  --- & $\Delta\emph{F814W} > 0.3$ mag \\
    Study: & this work & \cite{Huang_2020} & \cite{Huang18}  
\enddata
\label{tab:samples}
\tablecomments{A comparison of the criteria for the final Mira sample in M101 with those used for NGC 1559 and NGC 4258. As a result of differences in available data, we were unable to match the criteria exactly.}
\end{deluxetable*}

We retrieve all the WFC3/IR images from the Mikulski Archive for Space Telescopes (MAST) as {\tt flt} files. These are pipeline-proceesed, calibrated, flat-fielded individual exposures. We then generate drizzled and stacked images using {\tt Drizzlepac 3.1.8}, aligned relative to the ninth epoch (first new epoch) of observations. We create stacked images for each epoch and filter for time-series photometry and a deep detection image stacking all of the exposures for each filter. Each filter and epoch of observation contains at least two dithers, allowing us to sample the point-spread function (PSF) at a subpixel level and produce drizzled images with a scale of 0$\farcs$12/pixel, slightly below the native WFC3/IR scale of 0$\farcs$128/pixel.

We use the {\tt DAOPHOT/ALLSTAR/ALLFRAME} suite of crowded-field photometry software \citep{Stetson87, Stetson94}, to perform aperture and PSF photometry on the drizzled images. We generate a source list using the {\tt DAOPHOT} procedure {\tt FIND} to detect objects with $>$3$\sigma$ significance in the \emph{F160W} stacked image, taking into account the sky background, readout noise, and number of stacked images. We use the {\tt DAOPHOT} procedure {\tt PHOT} to perform aperture photometry on this source list, which is then input into {\tt ALLSTAR} to perform PSF photometry and generate a source-subtracted image. We then perform a second round of aperture and PSF photometry on the source-subtracted stacked image, resulting in a final source list of $\sim$100$,$000 point sources. 

In the next step, we input the source list into {\tt ALLFRAME} to perform time-series photometry on the 13 individual epochs. {\tt ALLFRAME} simultaneously fits the profiles of all sources across the full baseline of epochs, allowing it to maintain a constant source list through multiple epochs and improving image registration and photometry of sources closer to the detection limit. For \textit{F110W}, we transform the coordinates from the source list derived from the deeper \emph{F160W} stacked image and repeat these photometry steps to create the \textit{F110W} source list.

\subsection{Calibration}\label{sec:calibration}
We select 500 bright field stars to use for the zeropoint calibration. We use the ninth epoch as the reference observation for the calibration as we did for the photometry. We calculate the offsets between the reference and each of the other epochs. For each pair of observations, we calculate the differences in magnitude for each star and perform an iterative 3$\sigma$ clip to remove outliers before calculating the mean magnitude difference for that pair. We also 3$\sigma$ clip any problematic pairs that have a particularly large magnitude difference. In our case, we found that the mean offsets between the reference and each epoch were all quite small, ranging from 0.0008 mag to 0.01 mag, and we did not need to remove any epochs. We take the mean pairwise difference of all the epochs from the reference -- $0.0035$ mag -- to be the overall instrumental zeropoint offset. We then correct each epoch by the overall instrumental zeropoint as well as their individual offsets from the reference epoch. 

Next, we convert from PSF to aperture magnitudes using aperture corrections. This also corrects for imperfections in our PSF model. To determine the aperture corrections, we visually select bright and isolated stars using the stacked image. We then subtract all other detected sources from it and perform aperture photometry on the selected stars in the subtracted image. Then we calculate the difference between the PSF photometry and the aperture photometry ($\Delta m$) at a range of aperture sizes to derive a growth curve. Figure \ref{fig:growth_curve} shows the individual magnitude differences for each of the reference stars and the mean aperture correction. Finally, to correct to an infinite aperture, we use the WFC3/IR encircled energy tables\footnote{Tables available at \url{https://www.stsci.edu/hst/instrumentation/wfc3/data-analysis/photometric-calibration/ir-encircled-energy}} to determine the percentage of flux encircled at each aperture. We then add this to $\Delta m$, the aperture correction, to obtain the correction to an infinite aperture. Finally, we apply the Vega magnitude HST WFC3/IR photometric zeropoints \citep{Deustua_2017} of 26.042 mag for \emph{F110W} and 24.662 mag for F160W. To summarize, going from the instrumental magnitude $m_i$ to the final calibrated magnitudes $m_f$, 
\begin{equation}
    m_f = m_i + 2.5 \log(t) - a_{\rm DAO} + a_{\rm HST} + \Delta m + \Delta m_b
\end{equation}
where $t$ is the exposure time, $a_{\rm DAO}$ is 25, $a_{\rm HST}$ is 24.662 for F160W, and $\Delta m_b$ is the bias correction, which is discussed in detail later, in \S\ref{sec:crowding}. 

\section{Methodology} 
\label{sec:Methods}


We detect $\sim$100$,$000 point sources in the stacked image. From there, we derive a clean sample of Mira variables using cuts based on robustness of variability, amplitude, color, and period. The optical light curves of Miras are known to vary cycle-to-cycle in both shape and magnitude at a given phase \citep{Ludendorff_1928}, although the light curves of short-period, O-rich Miras are generally symmetric \citep{Vardya_1988} and often close to sinusoidal \citep{Lebzelter_2011}. At near-infrared and infrared wavelengths, their shapes are also well-approximated by a sinusoid. \cite{Huang18} found that cycle-to-cycle variations of the mean $H$-band magnitude are $\sim 0.07$ mag and that only about 1\% of stars with \textit{F160W} light curves required a higher-order Fourier component fit. Thus, following both \cite{Huang18} and \cite{Huang_2020}, we fit the periods and light curve parameters of our candidate Miras with a sine function. 

\begin{figure*}
\includegraphics[width = 0.49\textwidth]{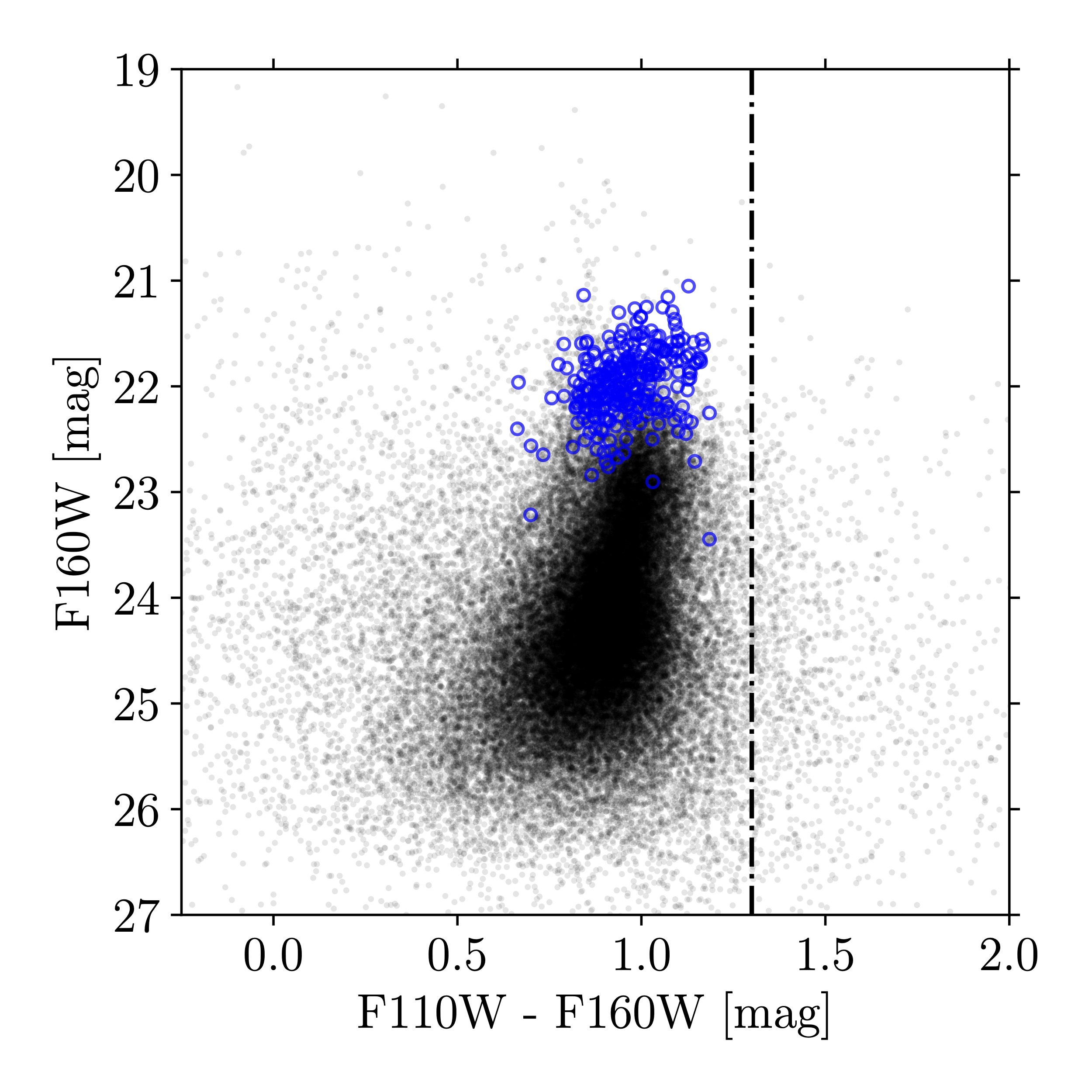}
  \includegraphics[width = 0.49\textwidth]{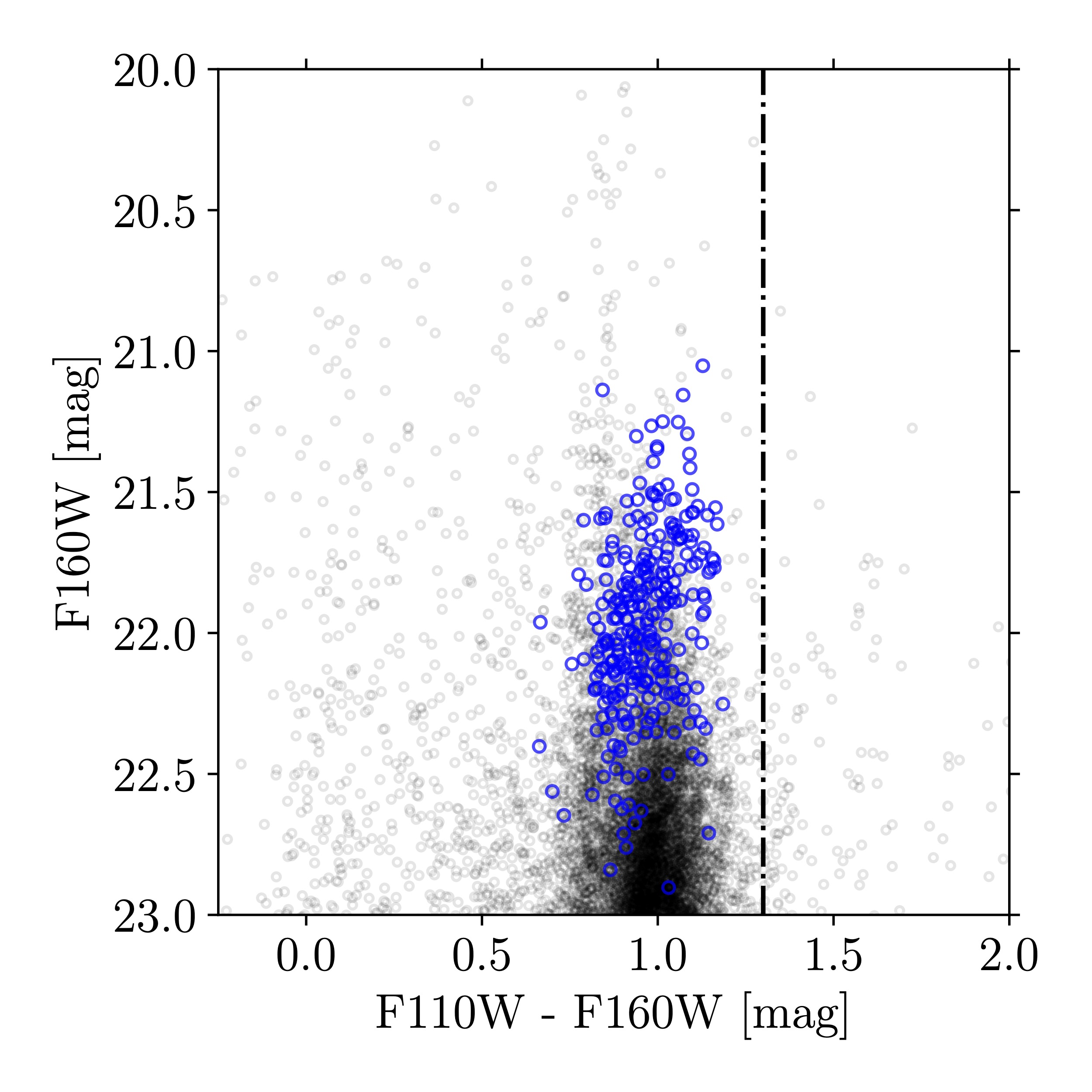}
\caption{\textit{Left:} Full \emph{F110W} - \emph{F160W}color magnitude diagram for M101 stars with the Miras are shown in blue. Dashed black line shows the color cut of $\emph{F110W} - \emph{F160W}< 1.3 $ mag. \textit{Right:} Same \emph{F110W}-\emph{F160W}color magnitude diagram but showing only the stars with magnitudes $\sim 0.5$ mag of the estimated Tip of the Red Giant Branch stars or brighter. 
\label{fig:color_mag}}
\end{figure*}

\subsection{Selection Criteria}
\label{sec:selection_criteria}

We use the {\tt TRIALP} code provided by Peter Stetson \citep{Stetson96} to obtain initial light curves and determine the Welch-Stetson variability index \textit{L} for the entire source list. This index selects for objects with a more sinusoidal or square-wave shape (rather than Gaussian variations, or random spikes) and also considers whether the variations are correlated between pairs of observations or pairs of filters. A lower $L$ value generally indicates that a particular source is less likely to be a {\it bona fide} variable. Conversely, a source with high $L$ value is more likely but not necessarily a variable. Thus, the $L$ index threshold is just a first step to create a starting list of possible variables which are then subject to other selection criteria. Figure \ref{fig:Lhist} shows the distribution of $L$ index values for all objects that had 10 or more epochs of observation. Similarly to our previous studies, we keep only sources with $L > 0.75$ to begin our search. 

After removing sources with fewer than 9 epochs of \textit{F160W} observation, we calculate periods for the remaining 3960 light curves using $\chi^2$ minimization on a grid search of periods ranging from 100 to 1000 days -- the relevant period range for Miras. Each light curve is fit with a sine function, leaving amplitude, mean magnitude, and phase as free parameters.

We then use an $F$-test 
to remove any sources whose fit does not favor a sinusoid over a first-order polynomial and apply an amplitude cut of $0.4$ mag $< \Delta \emph{F160W}< 0.8$ mag and an initial period cut of $200$ days $< P < 500$ days. Previously, in both \cite{Huang18} and \cite{Huang_2020}, we used $P < 400$d as an initial upper limit on period. This was limited both by the observational baseline ($\sim$400 days in both of the previous studies) which prevented us from measuring the periods of more slowly-varying stars and by physics (since longer-period Miras may be undergoing hot-bottom burning). However, thanks to the considerably longer observational baseline for M101, we are able to extend the initial search up to $500$ days. It is possible that some stars in the $400 - 500$ day range may be undergoing hot-bottom burning and thus bias the PLR. Thus, for the final distance measurement, both the upper and lower bounds of the period range are redetermined empirically in \S\ref{sec:completeness}. The minimum period is motivated by completeness while the upper bound is to avoid bright outliers which may be undergoing hot-bottom-burning. 

We next impose a color cut of $\emph{F110W} - \emph{F160W}< 1.3$. This removes the reddest objects, which are more likely to be C-rich. 
At this stage, we are left with 324 sources, which have passed all of the initial criteria for being Mira candidates. A summary of the cuts used and their comparisons with previous Mira searches in NGC 1559 and NGC 4258 can be found in Table \ref{tab:samples}.

\subsection{Bias Corrections}\label{sec:crowding}
In crowded-field photometry, measurement bias can result from a high density of unresolved background sources, which may affect the determination of the sky background and PSF fits. This typically has the effect of artificially increasing the brightness of measured sources. We correct for this bias due to crowding using artificial star tests, which are have also been employed in Cepheid distance measurements for decades \citep[e.g.,][]{2000PASP..112..177F, Hoffmann16, Yuan_2020, Riess22}. These crowding (or bias) corrections derived from the artificial star tests are a statistical correction. However, individual stars may also suffer from \textit{blending}, where a bright nearby source may make an individual Mira appear anomalously brighter and decrease its amplitude. Blending can be the result of physical association or close chance superposition but is generally not compensated by the crowding corrections from artificial star tests. Instead we consider Mira candidates which are within one full-width half maximum (2.5 pixels) of a source of the same magnitude or brighter to be blended and remove them from the sample (a total of 36 objects). 

For the remaining Mira candidates, we fit an initial PLR. To determine the starting magnitudes of the artificial stars, we use each star's period and the initial PLR to create an artificial star at the magnitude predicted by the PLR. The artificial star is then randomly dropped within a 20-pixel ($2.4''$) radius of the original Mira, so that both are in the same environment. To avoid introducing additional crowding into the image, we create 100 images, each including only one artificial star per Mira. We then perform aperture and PSF photometry on the images with injected artificial stars exactly as we would with real data. The difference in magnitude between the input artificial stars and the recovered sources is used to obtain an estimate of the crowding corrections. We then correct the magnitude of the original Miras using the crowding corrections and refit the PLR to determine a new input magnitude for the artificial stars. This process is repeated until the recovered magnitudes of the artificial stars converge with the measured magnitudes of the Miras. 

Overall, bias corrections for this galaxy were small (mean: 0.076 mag, median: 0.047 mag, $\sigma = $ 0.171 mag) but with a non-Gaussian distribution. The small size of the crowding corrections is likely due to M101's relative proximity compared to other supernova host galaxies and the more isolated SN~Ia field in which we have observed Miras compared to the more central disk regions targeted for Cepheids or Miras in other galaxies. Figure \ref{fig:crowding} shows that the distribution of crowding corrections for this galaxy is asymmetric with a long tail at the bright end. We interpret this as a real effect caused by the artificial stars occasionally landing close to a field star. While most lines of sight through the galaxy are clear, these occasional interlopers cause a long tail in the distribution of the crowding corrections towards the bright end. 

Rather than apply the median (biased estimator) or mean (unbiased estimator) of the final crowding corrections for each star, as done in \cite{Huang18} and \cite{Huang_2020} where the distribution of the corrections was nearly Gaussian, we instead account for the asymmetry by using a bootstrap Monte Carlo method to randomly draw a crowding correction for each Mira from its artificial star tests while assuming flat uncertainties for the crowding correction of each star. We then fit and 3$\sigma$ clip the corrected magnitudes with a PLR. The resulting distribution of the PLR zeropoints is approximately Gaussian, as expected from the central limit theorem. The distribution of zeropoints is shown in the right panel of Figure \ref{fig:crowding}. 
We take the mean and standard deviation of the resulting distribution as the zeropoint and the statistical uncertainty in the PLR when including the photometric uncertainties, intrinsic scatter, and uncertainties due to the crowding corrections. In \S\ref{sec:results}, we compare this result with that obtained from using the mean crowding correction for each star and fitting a PLR using linear regression and 3$\sigma$ clipping and find very good agreement between the two methods. 

\subsection{Refining the Period Range}
\label{sec:completeness}

C-rich Miras are the largest contaminant in the O-rich Mira PLR. However, they are also generally more massive and younger stars and are thus less common at shorter periods \citep{Feast_2006}. The mass at which C-rich asymptotic giant branch stars form is dependent on metallicity. At higher metallicities, C stars are rarer than O-rich AGB stars. This can be seen in the Milky Way, where the average period of a C-rich Mira is $P \sim 520$ d while in the LMC it is $\sim 400$ d. 
Thus, the period cut imposed on the short-period end is motivated by completeness, not astrophysics. The intrinsic width of the PLR can create a bias at shorter periods if the stars below the mean value of the PLR are not detected. 

At the upper range, we are both excluding C-rich stars and O-rich stars which may be undergoing hot-bottom burning. Similarly to \cite{Huang_2020}, we determine the limits empirically by using a boxcar with a width of 75 days to calculate the zeropoint of the PLR at a small range of periods. Figure \ref{fig:boxcar} shows the results of the boxcar fit. We find that the zeropoint is roughly flat when considering a range of starting periods between 260 and 400 days \citep[also relatively consistent with the range determined for NGC 1559 in][]{Huang_2020} but shows a trend at both short and long periods. At short periods, the zeropoint is biased bright --- consistent with completeness --- and at long periods, the zeropoint is biased faint, consistent with potential C-rich contamination. 

\subsection{Period-Luminosity Relation}
\label{sec:PLR}

In \cite{Huang_2020}, we compared the results from two slopes for a linear PLR. The first was determined for the $H$-band by \cite{Yuan17b} using ground-based data from OGLE-III and the Large Magellanic Cloud Near-infrared Synoptic Survey \citep{Macri15}. Since our observations are in \textit{F160W}, we also used a color transformation derived from O-rich Mira spectra to convert this into an \textit{F160W} slope (see \S3.4.3 of \cite{Huang18} for a more detailed explanation). The results for both slopes were found to be in $\sim 0.02$ mag agreement in the previous studies. Thus for simplicity, our best-fit result here uses only the PLR with the color-transformed \textit{F160W} slope,
\begin{align} \label{eqn:PLR}
m_{\rm F160W} &= a - 3.35 \cdot (\log P - 2.3)
\end{align}
where $m_{\rm F160W}$ is the magnitude in \textit{F160W} bandpass, $a$ is the zeropoint and magnitude of a 200-day Mira, and $P$ is the period in days. 

Total uncertainties for individual Miras ($\sigma_{\rm tot}$) are given by 
\begin{align}
\sigma_{\rm tot} = \sqrt{\sigma^2_{\rm int} + \sigma^2_{\rm phot}}
\end{align}
the quadrature sum of the photometric error ($\sigma_{\rm phot}, \sim 0.02 - 0.1$ mag) and the intrinsic scatter of the Mira PLR in the near-infrared ($\sigma_{\rm int}, \sim 0.13$ mag). We typically also include the uncertainty due to crowding bias in the total uncertainty for each Mira. However, in this case, the distribution of the individual crowding corrections is non-Gaussian, so the total uncertainty on the resulting zeropoint of the PLR comes from the results of the Monte Carlo bootstrap resampling and naturally incorporates the uncertainties from the crowding corrections already. 

\subsection{C-rich Contamination}
\label{sec:crich}

\begin{figure*}[]
\includegraphics[width=0.5\textwidth]{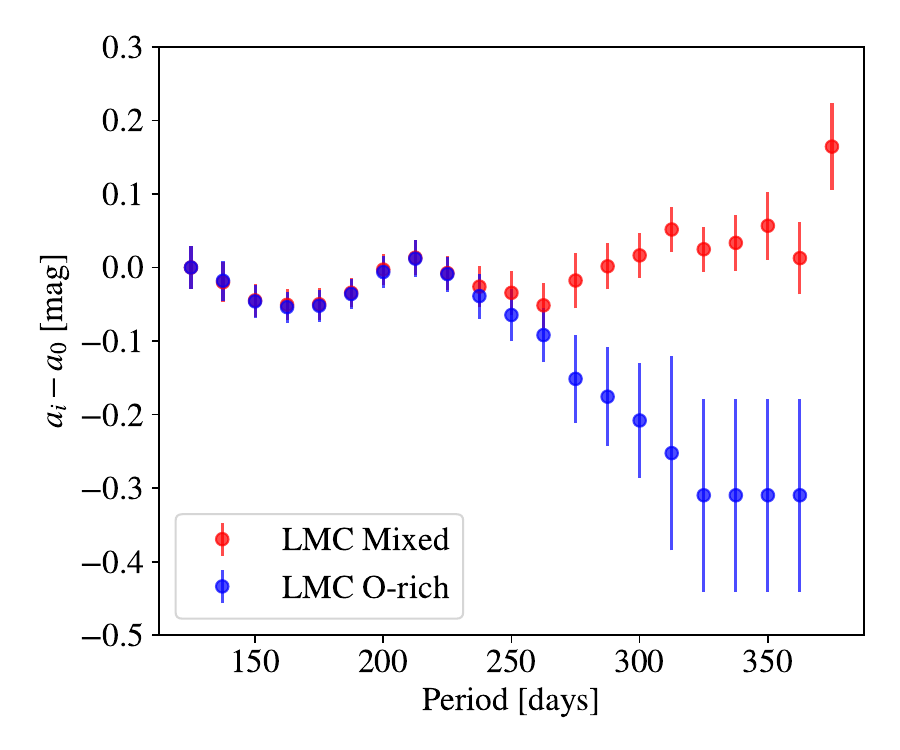}
\includegraphics[width = 0.5\textwidth]{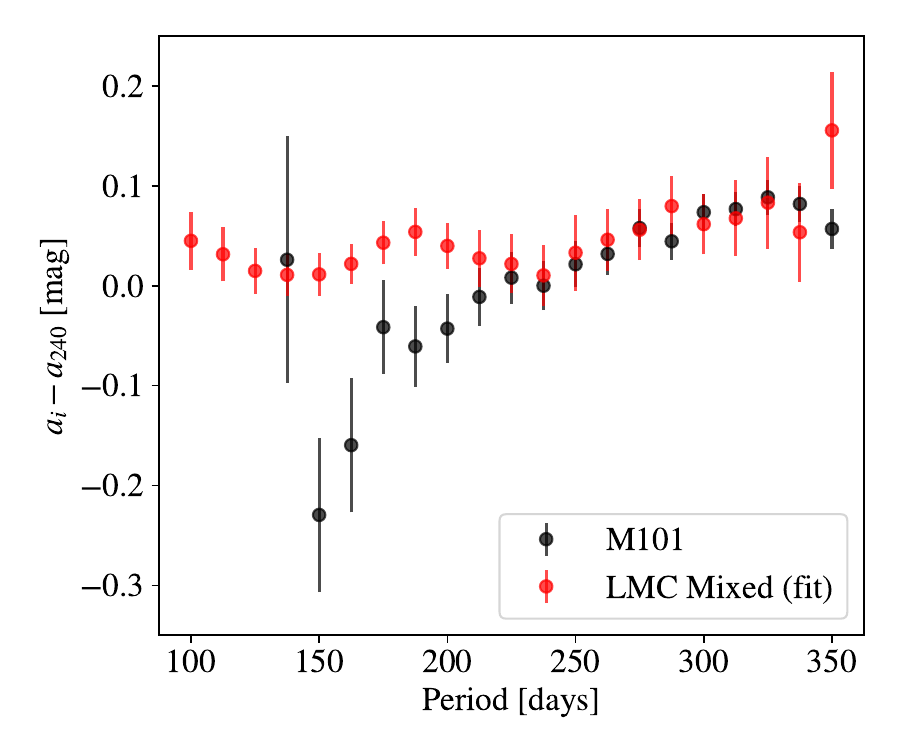}
\caption{{\bf Left:} The relationship between period and zeropoint for the two LMC populations of Miras we used in the the C-rich contamination model (spectrally-classified O-rich) and a mixed population. $a_i$ is the zeropoint at a particular period bin and $a_0$ is the zeropoint of the Mira PLR at 200 days. The mixed population consists of 33 C-rich Miras and 96 O-rich Miras that passed the period and amplitude cuts. {\bf Right:} The same mixed LMC population curve but fit using Equation \ref{eqn:crich} to the M101 points after excluding all of the bins below the completeness limit and the last bin.
\label{fig:crich}}
\end{figure*}

In ground-based studies, C- and O-rich Miras are often separated on the basis of their $J-H$ or $J-K_s$ colors. In these bands, C-rich AGB stars are typically redder than their O-rich counterparts. Similar to the cut used in \cite{Huang18} we apply a color cut of $\emph{F110W} - \emph{F160W}< 1.3$ to remove the reddest stars. Unfortunately, as we can see from Figure \ref{fig:color_mag}, the WFC3/IR color-magnitude diagram is relatively featureless and we cannot readily distinguish the spectral types of AGB stars in it. This is due to the NIR filter combination -- both bands are too broad to target the distinctive molecular features of the two spectral types which is in agreement with findings from other stellar populations. \cite{Dalcanton12b} showed that the majority of C-rich AGB stars in their stellar populations occupied similar locations in the WFC3/IR color-magnitude diagrams as the O-rich AGB stars. 

Since we cannot fully remove C-rich Miras from our samples, we instead estimate their potential bias on the PLR following the method introduced in \S3.5 of \cite{Huang_2020}. We use the $J$ and $H$-band observations of Miras from the LMC from \cite{Yuan17b} for comparison. We estimate the amplitudes of the variables using the minimum and maximum magnitudes reported in the dataset. The minimum and maximum $H$-band magnitudes in this dataset are sparsely sampled but cover multiple cycles. We then use the previously-derived color transformation for O-rich Miras from \cite{Huang18}, 
\begin{equation}
m_{F160W} = m_{H} + 0.39  \cdot (m_H - m_J)
\end{equation}
where $m_{F160W}$ is the magnitude in $F160W$, $m_H$ is the $H$-band magnitudes and $m_J$ is the $J$-band magnitude, to convert the ground-based $H$-band magnitudes and amplitudes to their $F160W$ equivalents. 


Next, we examine the difference in zeropoint for both of these samples as a function of period. Assuming that C-rich Miras will have a similar effect on the PLR in the LMC as in M101, we can correct for the change in the zeropoint due to C-rich Mira contamination by fitting the mixed zeropoint-vs-period curve from the LMC to the M101 zeropoint-vs-period relationship.  We calculate the zeropoint as a function of period in 50-day period bins for the contaminated sample of LMC Miras. Then, we model the zeropoint-vs-period curve of M101 with the LMC curve using, 
\begin{align}\label{eqn:crich}
Z_{\rm host}(P) = \alpha Z_{\rm LMC}(P) + \beta
\end{align}
where $Z_{\rm host}(P)$ is the zeropoint as a function of period for the Miras (assumed to be contaminated) in a given host. We then fit $\alpha$ and $\beta$, which simply scale and shift the zeropoints in the fit to match each galaxy. Scaling the size of the correction changes the level of C-rich contamination while shifting the zeropoint allows us to correct for the difference in the distance moduli of the hosts. We note that the difference between the zeropoint of the entire LMC mixed and O-rich samples in our relevant period range is $0.025$ mag, so this is a relatively small correction even for the C-rich dominated LMC.

For the fit, we only consider period bins that are above the completeness limit of M101 (estimated at 240 days). We find that for this range, the best fit is $\alpha = 0.19  \pm 0.16$. This small value of $\alpha$ is primarily driven by the final bin, which has the largest residual between the LMC and M101. Thus, if we refit while excluding this point we find $\alpha = 0.67 \pm 0.23$. However, neither of these fits is significantly better than when assuming no dependence (a straight line). As a result, we do not correct for C-rich contamination for the M101 sample. Figure \ref{fig:crich} shows the LMC curves for the mixed and pure O-rich populations as well as the fit of the LMC mixed curve to the M101 observations after excluding the longest-period bin and the periods below the completeness limit. While the fit appears good, the fluctuations in zeropoint are still within the uncertainties, and thus the curve is not significantly favored over the null result of no relation. This implies that the C-rich correction is likely not significant for a metal-rich galaxy like M101 (which will be dominated by O-rich stars) compared to a metal-poor galaxy like the LMC (which is dominated by C-rich stars). 

\section{Sources of Systematic Uncertainty} \label{sec:systematics}

\subsection{Extinction}\label{sec:extinction}

We correct for extinction from the Milky Way along our line of sight assuming an extinction to reddening ratio $A_V/E(B-V) = 3.1$ and using the $A_b/E(B-V)$ values for the WFC3 \textit{F110W} and WFC3 \textit{F160W} bands given in Table 6 of \cite{Schlafly11}. The differential extinction between the hosts and the anchors and the variation in extinction within the anchor and host galaxies can have an effect on the measurements. Using the dust maps, we correct for the extinction along the line-of-sight within the Milky Way. We find \emph{F160W} extinction values to NGC 4258, NGC 1559 and M101 of 0.008, 0.015, and 0.005 mag, respectively. 

While this addresses the extinction along the line-of-sight from the Milky Way, the \cite{Schlafly11} dust maps do not account for extinction within the Mira host galaxies. Furthermore, with only two colors, we are unable to estimate extinction to individual Miras using only our observations. Qualitatively, the Mira fields in the SN~Ia hosts are relatively homogeneous in their surface brightness/dust distribution and are typically not located near dusty regions of the galaxy. Extinction in the near-infrared is usually also relatively small. Thus, we expect internal extinction to be of similar scale to the Milky Way line-of-sight extinction and to be relatively consistent from star-to-star. Based on the consistency of the relative distance moduli from the LMC and NGC 4258 (which is more representative of a typical SN~Ia host), we estimate this internal extinction to be no more than $\sim 0.02$ mag. 

For Cepheids, the Wesenheit (``reddening free") $m^W_H$ mean magnitudes are often used to fit the PLR. While Wesenheit JK ($W_{JK}$) magnitudes appear to minimize the scatter of the Mira NIR PLR, there is evidence that a Mira PLR using Wesenheit JH ($W_{JH}$) mean magnitudes shows larger scatter than fitting a PLR using only the H-band mean magnitudes \citep[see][Figure 8]{Yuan17b}. Thus, without a $K$-band equivalent, we use only \emph{F160W} magnitudes for the PLR. 

\subsection{Metallicity}\label{sec:metallicity}

\begin{figure*}[ht!]
\plotone{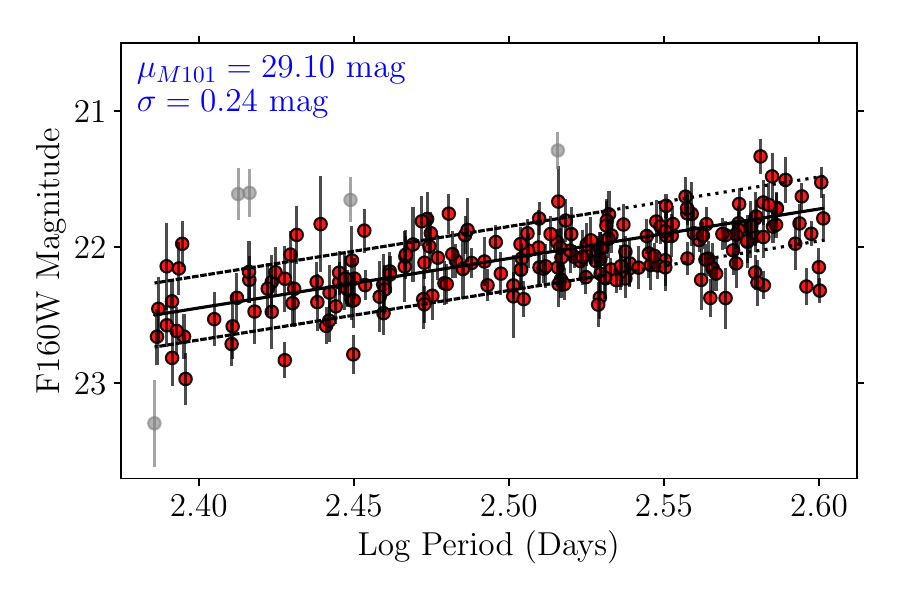}
\caption{Period-Luminosity relation with mean crowding correction of Miras $240 < P < 400$ days used in the baseline $H_0$ result. The red points are Miras used in the final fit and gray points are stars which ahve been removed through iterative 3$\sigma$ clipping. 
\label{fig:PLR}}
\end{figure*}

Theoretical models suggest that metallicity may have an effect on the Mira PLR \citep{Wood90, Qin18}. However, this effect has generally not been confirmed by observational studies. Since there has been no significant metallicity effect observed for Mira variables in the NIR, we only include a statistical uncertainty (of $\sim 0.03 $ mag) to account for such a potential undiscovered effect. This upper limit is estimated from the level of consistency between AGB star PLRs in environments of varying metallicity. \cite{Goldman19} found no discernible difference in the pulsational properties or the PLRs of stars in metal-poor (1.4\% solar metallicity) and more metal-rich (50\% solar) environments. Other studies have confirmed this when examining $K-$ band PLRs of stars in the LMC, SMC, Galaxy, and Galactic globular clusters \citep{Whitelock_1994, Whitelock_2008, Bhardwaj_2019}. 

Though we do not know the metallicity dependence of the Mira PLR, we can approximate the metallicity of the Mira field in M101 to confirm that it falls within the range of environments in which the pulsation of AGB stars have previously been studied. Using the relationship derived by \cite{Kennicutt_2003} from abundance measurements of HII regions in M101, the metallicity (or more precisely, the oxygen abundance) as a function of radial distance is given by
\begin{equation}
12 + \log ({\rm O}/{\rm H}) = 8.76(\pm 0.06) - 0.90(\pm 0.08)(R/R_0) 
\end{equation}
where $R$ is the radial distance from the center and $R_0$ is the scale length of the galaxy. Assuming an M101 distance of 7.5 Mpc, $R_0 = 32.4$ kpc. For the distances of the Mira field, this relation gives $12 + \log ({\rm O}/{\rm H}) \approx 8.4$ to $8.7$, somewhere between solar and LMC metallicity. This is well within the range of previously-studied AGB environmental metallicities, for which no discernible dependence was detected. Thus, it is unlikely that metallicity will have a measurable effect on the M101 distance. 

Environmental oxygen abundance is known to affect the ratio of O- to C-rich AGB stars. Stars entering the AGB with a lower photospheric C/O ratio require more carbon to be dredged to the surface to become C-rich (stars with C/O $> 1$ are considered C-rich and C/O $< 1$ are O-rich). Galaxies with higher metallicity typically exhibit larger ratios of O- to C-rich Miras \citep{Battinelli05, Hamren15, Boyer17}. Thus, cleanly separating spectral types or correcting for C-rich contamination is more important in lower metallicity environments where C-rich Miras may be more common. 

\section{Determining the Hubble Constant}
\label{sec:H0}

Determining $H_0$ is typically a three-step (or rung) process in which geometric calibrations to the more common primary standard candles such as Miras are used to determine the absolute magnitude of the P-L relation. Then, we use Miras to solve for the peak SN~Ia luminosity, $M_B^0$. Finally, SNe Ia in the Hubble flow (the regime where recessional velocities can be primarily attributed to the expansion of the universe and not local gravitational interactions) measure $H_0$ directly. 

\subsection{Anchors}
In the first rung, we use the same anchors as \cite{Huang_2020} -- the water megamasers in NGC 4258 with distances determined in \cite{Pesce_2020} and the detached eclipsing binary distance to the LMC from \cite{Pietrzynski19} which are precise to 1.5\% and 1.2\% respectively. The geometric distance to NGC 4258 is $\mu_{N4258} = 29.397 \pm 0.0324$  mag and the geometric distance to the LMC is $\mu_{LMC} = 18.477 \pm 0.004$ (stat) $\pm 0.026$ (sys) mag. A new set of WFC3/IR photometric zeropoint calibrations has been published since \cite{Huang_2020} (previously the infinite aperture Vegamag zeropoint for \emph{F160W} was 24.6949 mag and it is currently 24.662 mag). Solving for $a_0$, the absolute magnitude of the Mira P-L zeropoint, while accounting for the new Vegamag zeropoints, we find that $a_0 = -6.30 \pm 0.053$ mag using the LMC and  $a_0 = -6.28 \pm 0.043$ mag for NGC 4258. 

For the final Mira P-L relation zeropoint, we take the weighted mean of these two calibrations, which gives 
\begin{align} \label{eqn:zpt}
a_0 = -6.29 \pm 0.033 {\rm \ mag}
\end{align}
using inverse-variance weighting. 

The Milky Way has been suggested as a potential third anchor for the Mira distance ladder by \cite{Sanders_2023}. They derived a calibration of the Mira PLR using candidates from the \textit{Gaia} long-period variable catalogue \citep{Mowlavi_2018}, single epoch NIR photometry from 2MASS, and parallaxes from \textit{Gaia}. Their final results have a steeper slope than previous studies, which causes the variables to be fainter than the previous LMC calibrations at the short-period end. 

However, several other works have shown that even with \textit{Gaia} DR3, obtaining precise parallax distances to Miras and other very red AGB stars remains challenging \citep{ElBadry_2021, Andriantsaralaza_2022, Maiz_Apellaniz_2022}. \cite{ElBadry_2021} used the nearly-identical parallaxes of the binary companions to verify \textit{Gaia's} parallax measurements between pairs of stars, demonstrating that bright red stars (which include AGB stars such as Miras) with well-behaved uncertainties can still have parallax uncertainties underestimated by $\sim 30-80\%$. \cite{Andriantsaralaza_2022} confirmed this result when comparing \textit{Gaia} parallaxes with very long baseline interferometry (VLBI) parallaxes of AGB stars with masing circumstellar envelopes. They showed empirically that parallax uncertainties for AGB stars are greatly underestimated, in agreement with theoretical models \citep{Chiavassa_2018}. \cite{Andriantsaralaza_2022} also found significant, asymmetrical errors for over 40\% of their sample of AGB stars. While VLBI parallaxes (such as those in \citealt{Andriantsaralaza_2022}) using maser emission in the circumstellar envelope of AGB stars can provide accurate parallaxes to nearby AGB stars, the subset of Miras with circumstellar maser emission is different from those that we used in the distance ladder. 

These large uncertainties are expected for \textit{Gaia} DR3 parallaxes --- Galactic Miras exist at a confluence of several difficulties for \textit{Gaia}. They are bright, very red, extended objects with changing chromaticity, all of which contribute to parallax errors and/or uncertainties. Miras have large convection cells, and many have radii $>1$ AU. As parallax and radius (angular size) both scale as $\sim 1/d$, this means that Miras for which parallaxes can be obtained will also be resolved and convection will cause their photocenters to move \citep[e.g.][]{Chiavassa_2018}. 

While the Milky Way may eventually serve as a reliable third anchor for the Mira distance ladder, given the disagreement with previous literature findings, the large scatter in the Mira PLR from \textit{Gaia} \citep[as evidenced by Figure 5 of][]{Sanders_2023}, and large individual uncertainties of objects fit in the \textit{Gaia} calibration, we currently use only NGC 4258 and the LMC as anchors in our analysis. 


\subsection{SN~Ia Calibrators}
\label{sec:sniacalibrators}

In the second rung, we compare the apparent magnitude zeropoint of the Mira P-L relations, $a_{\rm host}$, with the apparent standardized SN~Ia magnitudes, $m_B^0$, in the same galaxies. $m_B^0$ is the maximum-light apparent magnitude of the SN~Ia which has been corrected for variations of fiducial color, luminosity, and host dependence in accordance with the guidelines of Pantheon+ \citep{Scolnic_2022, Brout_2022a}. Similar to \cite{Scolnic23}, we also define the difference in apparent magnitude for the two standard candles (Miras and SN~Ia) in the same host, $\Delta S$, which is
\begin{align}\label{eqn:DeltaS}
\Delta S = a_{\rm host} - m_B^0.
\end{align}
We can then calculate $\overline{\Delta S}$, the weighted average of $\Delta S$ over the hosts with the weights again given by inverse-variance weighting.  This is the weighted average of difference in apparent magnitude for individual hosts using the sum of the inverse error squared for each component's weight. In \cite{Huang_2020}, the single calibrating SN~Ia was SN 2005df, with $m^0_B = 12.141 \pm 0.086$ mag \citep{Scolnic_2022}.  Again updating this with the new Vegamag zeropoint for \textit{F160W}, and using Miras with $260\!<\!P\!<\!400$\ days. This was the completeness range for NGC 1559 (the host of SN 2005df) determined via the boxcar fit and used in the baseline distance determination to NGC 1559. With this period range of Miras, we find $a_{\rm N1559} = \ $\aN \ mag, which gives $\Delta S = 12.966 \pm 0.107$\ mag from this galaxy.  

From \cite{Scolnic23}, the color and light-curve corrected peak magnitude of the SN~Ia in M101, SN 2011fe, is $9.808 \pm 0.116$ mag. From the Miras with $240\!<\!P\!<\!400$\ days in M101 (again with period range estimated using the boxcar fits), we determine $a_{\rm M101} = \ $\aM \ mag and $\Delta S = $ \DSMessier \ mag. The final PLR determined for M101 is shown in Figure \ref{fig:PLR}. 
Taking the weighted mean of these two values gives us $\overline{\Delta S} =$ \DSavg \ mag. This $\overline{\Delta S}$ is the mean difference in brightness between the peak luminosity of a SN~Ia and the mean magnitude of a 200-day Mira. 

The geometric calibration from the anchors and the magnitude difference in the SN~Ia hosts yield $M_B^0$, the fiducial peak SN~Ia luminosity, as follows
\begin{align}\label{eqn:MB0}
 M_B^0 = a_0 - \overline{\Delta S}    
\end{align}
\noindent from which we obtain $M_B^0 = $\snfemb \ mag. 


\subsection{Hubble Flow SNe}
\label{sec:thirdrung}

In the third rung, SNe Ia in the ``Hubble flow"--- where their recessional velocities are dominated by the expansion of the Universe rather than by local peculiar gravitational interactions --- are used to measure $a_B$, the intercept of the magnitude (or distance) and redshift relation (Hubble diagram). Thus, this rung is completely separable from the previous two and does not directly involve Miras. For an arbitrary expansion history and redshift, $a_B$ is given by
\begin{equation}
\begin{aligned}
    a_B &= \log cz \bigg[1 + \frac{1}{2} \left(1- q_0\right)z  \\
    &- \frac{1}{6}\left(1 - q_0 - 3q_0^2+ j_0 \right)z^2 + \mathcal{O}(z^3) \bigg] - \frac{1}{5}m^0_B,
\end{aligned} \end{equation} where $q_0$ is known as the deceleration parameter, and $j_0$ is the jerk --- the second and third order time derivatives of the scale factor, respectively \citep{Visser_2004}. For measuring $H_0$, we consider the small redshift limit ($z \approx 0$). This equation then approximates to, 
\begin{align}
a_B = \log cz - \frac{m_B^0}{5}
\end{align}
where $z$ is redshift and includes peculiar velocity corrections. 

\setlength{\tabcolsep}{1em}
\begin{deluxetable*}{l|l|cc|cc}
\tabletypesize{\scriptsize}
\tablecaption{Sources of Uncertainty}
\tablewidth{0pt}
\tablehead{ \multirow{2}{*}{Term} & \multirow{2}{*}{Description} & \multicolumn{2}{c|}{Huang+ (2020)} & \multicolumn{2}{c}{This work} \\
 & & LMC & NGC 4258 & LMC & NGC 4258}
\startdata
   $\sigma_{\mu,{\rm anchor}}$ [mag] & Anchor distance &  0.0263 &  0.032 & 0.0263 & 0.0324 \\
   $\sigma_{{\rm PLR,anchor}}$ [mag] & Mean of the PLR in the anchor galaxy & 0.010 & 0.017 & 0.010 & 0.017 \\
   $\sigma_{\rm anchor}$ [mag] & Total anchor uncertainty & \multicolumn{2}{c|}{0.022} & \multicolumn{2}{c}{0.022} \\
   \hline
    $\sigma_{\rm PLR}$ [mag] & PLR slope, differential extinction, metallicity & \multicolumn{2}{c|}{0.047} & \multicolumn{2}{c}{0.047} \\
   $\sigma_{\rm PL, hosts}$ [mag] & Mean of PLR in SN~Ia host(s) &  \multicolumn{2}{c|}{0.038} & \multicolumn{2}{c}{0.015} \\
   $\sigma_{\rm SN}$ [mag]& Mean of SN~Ia calibrator(s) (\# SN) & \multicolumn{2}{c|}{0.11 (1)} & \multicolumn{2}{c}{0.069 (2)}\\
   $\sigma_{aB}$ [mag] & Intercept of SN~Ia Hubble diagram & \multicolumn{2}{c|}{0.00176} & \multicolumn{2}{c}{0.0085} \\
   \hline
   \hline
    \multicolumn{2}{l|}{Total Systematic Uncertainty [mag]} & \multicolumn{2}{c|}{0.045} & \multicolumn{2}{c}{0.045}  \\
    \hline
    \multicolumn{2}{l|}{Total Statistical Uncertainty [mag]} & \multicolumn{2}{c|}{0.116} & \multicolumn{2}{c}{0.074} \\
   \hline
   \multicolumn{2}{l|}{Total uncertainty on $\sigma_{H_0}$ [\%]} &  \multicolumn{2}{c|}{5.5} & \multicolumn{2}{c}{4.1}
\enddata
\label{tab:uncertainties}
\tablecomments{The error budget for our measurement compared with the error budget for the previous Mira-based $H_0$ from \cite{Huang_2020}. Numbers are approximations for comparison since the two papers follow two slightly different procedures (\cite{Huang_2020} did not use Monte Carlo bootstrapping or $\Delta S$ formulation) for determining $H_0$. \\}
\end{deluxetable*}

In our determination of $H_0$ we use the baseline value of $a_B$ from \cite{Riess22}, 
\begin{align}
a_B = 0.714158 \pm 0.0085
\end{align}
which is determined from the Hubble diagram of 277 SNe Ia in the Hubble flow. These are supernovae from the Pantheon+ sample \citep{Scolnic_2022, Brout_2022a} which have have $0.0233 < z < 0.15$, pass the same quality cuts, and share the properties of Cepheid hosts, which are in late-type galaxies \citep{Riess22}. Despite the fact that Miras are a ubiquitous older population, this value for $a_B$ is appropriate for our determination since the two local Mira calibrators (NGC 1559 and M101) are both also Cepheid hosts. However, if the sample of Mira-SN~Ia hosts is eventually expanded to include early-type hosts as well then it may be more appropriate to include an $a_B$ value determined using Hubble flow SNe Ia in all host types. 

\subsection{Sources of Uncertainty}\label{sec:Uncertainties}

Table \ref{tab:uncertainties} contains a summary of the various sources of uncertainty in our measurement in comparison to the previous Mira-SN $H_0$ measurement from \cite{Huang_2020}. The primary reduction in uncertainty originates from the addition of a second SN~Ia calibrator. In addition, the constraint on the peak magnitude of SN 2005df (the SN~Ia in NGC 1559) has improved. \cite{Huang_2020} used the Pantheon+ calibration of $m_B = 12.14 \pm 0.11$ mag from \cite{Scolnic18}. In the new Pantheon+ SN~Ia sample, the peak magnitude for this supernova is now $m_B = 12.141 \pm 0.086$ mag. 

The systematic uncertainties primarily originate from the uncertainty in the slope of the PLR (which is relatively small, at $\sim$ 0.01 mag) and the  uncertainty in the differential extinction and metallicity between the anchors and the hosts.  Unlike NGC 4258 and the SN~Ia host galaxies, which are late-type spiral galaxies, the LMC is an irregular dwarf galaxy. In \cite{Huang18} we investigated the potential effects of the different environments by comparing the relative distance modulus of the LMC and NGC 4258 determined using Miras with the difference in their distance moduli determined using geometric methods. We found no indication of a systematic difference and that these two relative distances were consistent within their statistical uncertainties. 

This agrees well with the assumption that internal (to the host or anchor galaxy) extinction is low in the near-infrared in the locations of short-period Miras and that there is little dependence of the PLR on metallicity. Therefore our systematic uncertainty budget is relatively conservative. 

\setlength{\tabcolsep}{1em}
\begin{deluxetable*}{clllllcccc}
\tabletypesize{\scriptsize}
\tablecaption{Final Mira Sample}
\tablewidth{0pt}
\tablehead{ Star ID & \colhead{$\alpha$} & \colhead{$\delta$}  & \colhead{Period} & \colhead{X} & \colhead{Y} & \colhead{$F160W$} & \colhead{$\sigma_{F160W}$} & \colhead{Amp} & \colhead{$\Delta m_b$} \\
 &  \multicolumn{1}{c}{(J2000)} & \multicolumn{1}{c}{(J2000)}  & \multicolumn{1}{c}{[days]} &  \multicolumn{1}{c}{[pix]} & \multicolumn{1}{c}{[pix]} & \multicolumn{1}{c}{[mag]} & \multicolumn{1}{c}{[mag]} & \multicolumn{1}{c}{[mag]} & \multicolumn{1}{c}{[mag]} }
\startdata
     11847 & 14$^{\rm h} 3^{\rm m} 8.837^{\rm s}$ & $+54^{\circ}17'13.21''$ &  278.797 &  955.520 &  343.315 &   22.202 &    0.023 &    0.518 &    0.055 \\ 
     15574 & 14 3 07.127 & +54 17 03.79 &  375.028 &  839.670 &  434.390 &   21.745 &    0.039 &    0.784 &    0.079 \\ 
     17867 & 14 3 10.547 & +54 16 54.12 &  327.769 & 1096.264 &  488.428 &   21.906 &    0.017 &    0.489 &    0.056 \\ 
     19646 & 14 3 07.606 & +54 16 51.86 &  323.505 &  884.815 &  529.661 &   22.125 &    0.031 &    0.469 &    0.023 \\ 
     20101 & 14 3 04.121 & +54 16 53.66 &  301.450 &  630.333 &  541.294 &   22.212 &    0.026 &    0.703 &    0.053 \\ 
     20855 & 14 3 10.783 & +54 16 45.39 &  339.800 & 1121.022 &  558.962 &   22.190 &    0.026 &    0.760 &    0.038 \\ 
     23593 & 14 3 08.969 & +54 16 39.23 &  375.068 &  994.737 &  623.863 &   21.824 &    0.038 &    0.619 &    0.048 \\ 
     23699 & 14 3 00.713 & +54 16 46.39 &  269.624 &  389.271 &  627.473 &   22.101 &    0.040 &    0.805 &    0.205 \\ 
     23734 & 14 3 00.420 & +54 16 46.58 &  328.050 &  367.864 &  628.169 &   21.572 &    0.049 &    0.561 &    0.092 \\ 
     24253 & 14 3 04.362 & +54 16 41.55 &  344.309 &  658.342 &  639.815 &   21.787 &    0.021 &    0.664 &    0.046 \\ 
     24555 & 14 3 10.021 & +54 16 35.52 &  278.777 & 1074.337 &  646.583 &   22.124 &    0.018 &    0.455 &    0.063 \\ 
\enddata
\label{tab:Mirasample}
\tablecomments{Miras used in the distance measurement to M101. \textit{F160W} magnitudes are calibrated and in Vegamag but do not include the crowding corrections. $\sigma_{F160W}$ includes only the photometric uncertainties (when fitting we also include an intrinsic PLR uncertainty of $\sim 0.12$ mag). $\Delta m_b$ is the mean crowding correction for each Mira derived from artificial star tests. The table is available in its entirety in a machine-readable format.}
\end{deluxetable*}

\begin{figure*}
\plotone{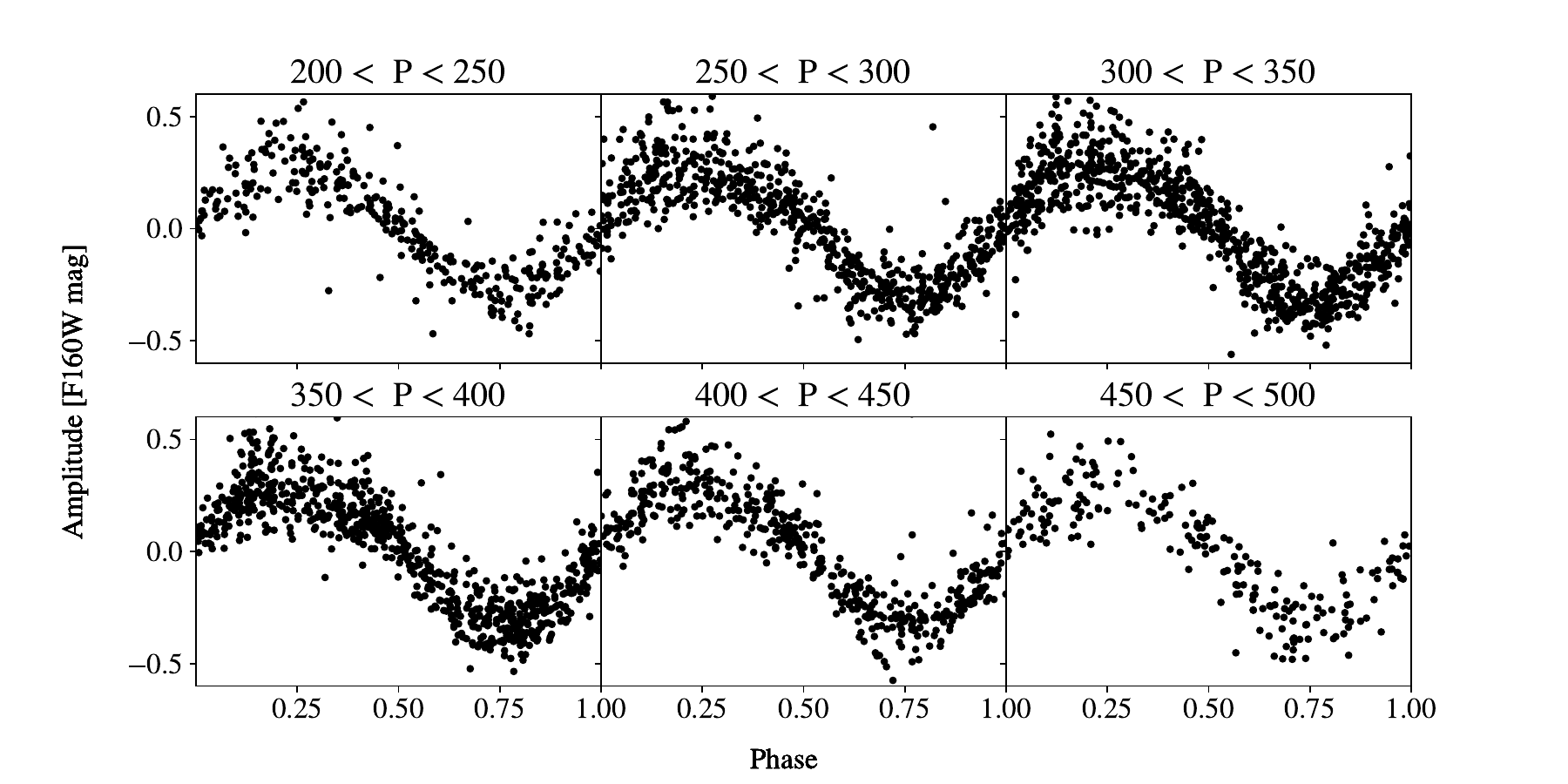}
\caption{The composite, phased light curves of Miras at a range of periods in our sample and exhibiting a roughly sinusoidal variation. Mean magnitudes have been subtracted and individual observations converted to their fit phase, but amplitudes of the stacked Miras have not been scaled.} 
\label{fig:stacked_lcs}
\end{figure*}

\section{Results and Discussion}\label{sec:results}

\subsection{The Final Mira Sample}

Our final sample consists of 211 Miras with periods ranging from $200-500$ days. Their properties are listed in Table~\ref{tab:Mirasample}. Figure \ref{fig:stacked_lcs} shows the composite light curve of all of the Miras plotted as a function of phase for a range of period bins. These light curves have not been scaled by amplitude and is thus also representative of the spread in amplitudes for our sample. We find that the light curves are approximately sinusoidal in \textit{F160W}. In addition, this allows us to verify the phasing and period determinations since the observations span between 6-15 cycles depending on the period of each Mira. 

\subsection{The Mira Distance to M101}

Using Equations \ref{eqn:PLR} and \ref{eqn:zpt}, and applying a bootstrap Monte-Carlo fit of the crowding corrections as described in \S\ref{sec:crowding}, we determine the distance modulus to M101 of
\begin{align}\label{eqn:mu}
\mu_{\rm M101} = 29.10 \pm 0.06 {\rm \ mag}
\end{align}
which includes systematic and statistical uncertainties. If we instead fit for the distance modulus similar to \cite{Huang_2020} by using linear regression, mean crowding corrections, and 3$\sigma$ clipping where we remove one outlier at a time, we find that the distance modulus is $\mu_{\rm M101} = 29.10 \pm 0.06 {\rm \ mag}$ which is identical to bootstrapping result. The fit for this result is shown in Figure \ref{fig:PLR}. 

\begin{figure*}
\plotone{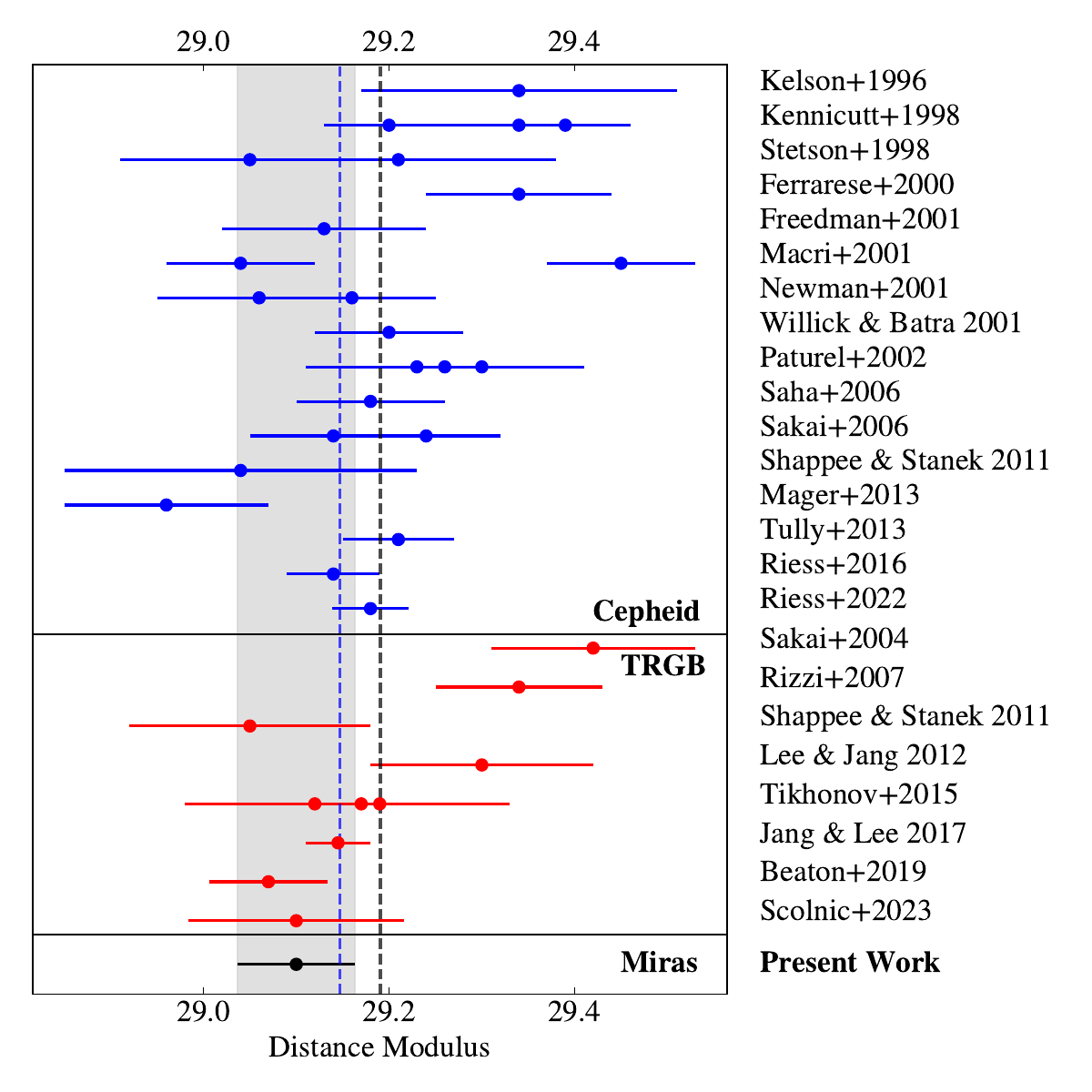}
\caption{Literature distance moduli to M101 measured using Cepheid (blue) and TRGB (red). Gray shaded region and black point shows the measurement from this work (Miras) of $\mu_{\rm M101} =$ \mugal \ mag. The weighted mean distance modulus of all of the previously published measurements is $\mu_{\rm M101} = 29.19 \pm 0.01$ mag (black dashed line) and the weighted mean distance modulus of the measurements published in the past 15 years for all calibrators is $\mu_{\rm M101} = 29.15 \pm 0.02$ mag (blue dashed line) and falls within the 1$\sigma$ of the distance derived here using Miras.  
\label{fig:comparisons}}
\end{figure*}

While this is the first Mira distance to this galaxy, there have been numerous previous distance measurements to M101 using a variety of other methods and distance indicators. Here, we focus on comparisons with literature measurements made using Cepheids and TRGB since these are the intermediate distance indicators most commonly used in the extragalactic distance ladder.  Table \ref{tab:comparisons} summarizes the literature distances determined to M101 using these indicators. A comparison of our result and these literature distances is shown in Figure \ref{fig:comparisons} along with the weighted (by the uncertainty) mean of all of the TRGB and Cepheid distance moduli to this galaxy ($\mu_{\rm M101} = 29.19 \pm 0.01$ mag) and the weighted mean Cepheid and TRGB distance modulus when using only measurements from the past 15 years ($\mu_{\rm M101} = 29.15 \pm 0.02$ mag). Our results are within 1$\sigma$ from the mean of the recent distance moduli measured for this galaxy, and in 2$\sigma$ agreement with nearly all of the recent measurements.  The agreement with literature distance moduli is even better when taking an unweighted average of the recent measurements, which gives $\mu_{\rm M101} = 29.13 \pm 0.02$ mag. 

Since there are no previous Mira distances to the galaxy, in Table \ref{tab:comparisons} and Figure \ref{fig:comparisons} we have used the final TRGB and Cepheid distances derived by the authors without attempting standardization. However, a direct comparison of the TRGB apparent magnitudes (which can eliminate some potential zeropoint calibration differences) can be found in \cite{Beaton_2019}. We also refer the reader to \cite{Beaton_2019} for detailed review and context of TRGB and Cepheid (Leavitt Law) distances published before 2019. Since \cite{Beaton_2019}, there has also been one new Cepheid distance \citep{Riess22} and one new TRGB distance \citep{Scolnic23} which we will briefly discuss here.

\setlength{\tabcolsep}{1em}
\begin{deluxetable*}{lll}
\tabletypesize{\small}
\tablecaption{Measured Distances to M101}
\tablewidth{0pt}
\tablehead{\multicolumn{1}{l}{Reference} & \multicolumn{1}{l}{Distance Modulus (mag) } & \multicolumn{1}{l}{Notes}
}
\startdata
\textbf{Cepheid Distances} \\
\hline 
    \cite{Kelson_1996} & $29.34 \pm 0.17$ \\
    \cite{Stetson98} & $29.05 \pm 0.14$  \\
    & $29.21 \pm 0.17 $  \\
    \cite{Kennicutt_1998} & $29.20 \pm 0.07$ & \\
    & $29.34 \pm 0.08$ & \\
    & $29.39 \pm 0.07$ & \\
   \cite{Ferrarese_2000} & $29.34 \pm 0.10$  \\
   \cite{Macri_2001} & $ 29.04 \pm 0.08$ & Inner field, {\emph F160W}\\
   & $29.45 \pm 0.08 $ & Outer field, {\emph F160W}\\
   \cite{Newman_2001} & $29.06 \pm 0.11$  & \\ 
   & $29.16 \pm 0.09$ & \\
   \cite{Willick_and_Batra_2001} & $29.20 \pm 0.08$ & \\
   \cite{Freedman01} & $29.13 \pm 0.11 $ & final result of the HST Key Project \\
   \cite{Paturel_2002} & $29.30 \pm 0.07$ & \\
   & $29.23 \pm 0.07 $ \\
   & $29.26 \pm 0.15 $\\
   \cite{Sakai_2004} & $29.14 \pm 0.09$ & \\
   & $29.24 \pm 0.08$ & \\
   \cite{Saha_2006} & $29.18 \pm 0.08$ & \\
    \cite{Shappee_2011} & $29.04 \pm 0.05$ (stat) $\pm 0.18$ (sys)  & \\
    \cite{Mager_2013}  & $28.96 \pm 0.11 $\\ 
     \cite{Tully_2013} & $29.21 \pm 0.06 $ \\ 
    \cite{Nataf_2015} & $29.20 \pm 0.03$ & Using Cepheid sample from \cite{Shappee_2011}\\
    \cite{Riess16} & $29.14 \pm 0.05$ & SH0ES 2016 result \\
    \cite{Riess22} & $ 29.178 \pm 0.041$  & Distance without inclusion of SN, SH0ES 2022 result \\
    \hline
    \textbf{TRGB Distances}\\
    \hline
    \cite{Sakai_2004} & $29.42 \pm 0.11$ & \\
    \cite{Rizzi_2007} & $29.34 \pm 0.09$ & \\ 
    \cite{Shappee_2011} & $29.05 \pm 0.06 \ {\rm (stat)} \pm 0.12 \ {\rm (sys)} $ & \\
    \cite{Lee_and_Jang_2012} & $29.30 \pm 0.01$ (stat) $\pm 0.12$ (sys)  \\
    \cite{Tikhonov_2015} & $29.12 \pm 0.14$ & \\
    & $29.17 \pm 0.13$ & \\
    & $29.19 \pm 0.14$ & \\
    \cite{Jang_and_Lee_2017} & $29.145 \pm 0.035$ \\ 
    \cite{Beaton_2019} & $29.07 \pm 0.04$ (stat) $\pm 0.05$ (sys) & Carnegie-Chicago Hubble Program result \\
    \cite{Scolnic23} & $29.10 \pm 0.116$ & Assuming tip luminosity $M^{R=4}_{I,{\rm TRGB}} = -4.030 \pm 0.035$ mag \\
    \hline
    \textbf{Mira Distances} \\
    \hline
    Huang et al. 2023 (this work) & \mugal  
\enddata
\tablecomments{Distances published before 2019 are compiled from the NASA/IPAC Extragalactic Database.}
\label{tab:comparisons}
\end{deluxetable*}

\cite{Riess22} fit a Leavitt Law using a total of 259 Cepheids found using observations in two M101 fields (called Field 1 and Field 2) on opposite sides of the nucleus of the galaxy. The original observing campaign was carried out in 2006 and each field was visited 12 times with a temporal spacing of $\sim$1-5 days between visits.  \cite{Riess22} concluded that even with addition of two later epochs of observation which were spaced by one week, the overall baseline was insufficient to provide reliable periods for Cepheids with $P > 35$ days. Therefore, they excluded M101 Cepheids with $P > 35$ days from their analysis, approximately 10\% of the \cite{Riess16} Cepheid sample for this galaxy. M101 was also one of the two nearest SN~Ia hosts, so the mean Cepheid period for this galaxy was shorter, at 15.8 days, compared to an average of 36.5 days across the entire SN~Ia calibrator sample. The Cepheid-calibrated distance to this galaxy that they derive without the inclusion of any SNe Ia is $\mu_{\rm M101} = 29.178 \pm 0.041$ mag and is in relatively good agreement with our results, with about a 1.2$\sigma$ difference. The standard process for detecting and classifying the Cepheids is explained in greater detail in \cite{Hoffmann16} and \cite{Riess22}.

The second recent measurement from \cite{Scolnic23} uses an unsupervised algorithm called Comparative Analysis of TRGBs (CATs) to reduce variance in the distances derived from different TRGB halo fields. They define a quantity \textit{R}, known as the contrast ratio, which is the ratio of the number of stars 0.5 magnitudes below vs. above the tip. For M101, They reported a raw (uncorrected for extinction) $I$-band TRGB magnitude of $m_{I, {\rm TRGB}} = 25.080 \pm 0.111$ mag and $R=4.4$. Using their Equation 1, this can be standardized to give $m^{ R{\rm=4}}_{I,{\rm TRGB}} = 25.072 \pm 0.111$ mag. 
Their absolute geometric calibration from NGC 4258 gives a fiducial tip luminosity of $M^{ R{\rm=4}}_{I,{\rm TRGB}} = -4.030 \pm 0.035$ mag. The resulting distance modulus obtained with this combination of apparent magnitude and tip luminosity is $\mu_{\rm M101} = 29.10 \pm 0.115$ mag, which is nearly identical to the Mira distance to this galaxy. 

\subsection{Hubble Constant Measurement}

With the peak SN~Ia magnitude determined in \S\ref{sec:sniacalibrators} from the combination of the first two rungs and the $a_B$ determined from the Hubble diagram of SN~Ia as described in \S\ref{sec:thirdrung}, we can solve for $H_0$ using, 
\begin{align}
    \log H_0 = (M_B^0 + 5a_B + 25)/5
\end{align}
which gives us 
\begin{equation}
    H_0 =  72.37 \pm 2.97 { \ \rm km \ s}^{-1} {\ \rm Mpc.}^{-1}
\end{equation}
This result includes both systematic (discussed in \S\ref{sec:systematics}) and statistical uncertainties and is a 4.1\% measurement of $H_0$. Like the previous Mira-based $H_0$ measurement from \cite{Huang_2020}, the uncertainty remains dominated by the statistical uncertainty in the peak magnitude of the SN~Ia calibrators. However, the addition of a second SN~Ia calibrator, has decreased the overall error budget by $\sim 1/\sqrt{2}$ from 5.5$\%$ to 4.1$\%$. 

This result is in very good agreement with the most recent Cepheid measurement from the SH0ES team \citep{Riess22} which has a baseline result of $73.04 \pm 1.04$ km s$^{-1}$ Mpc$^{-1}$.  More relevant is the Cepheid result that only uses the same anchors as we use, NGC 4258 and LMC, $73.35 \pm 1.17$ km s$^{-1}$ Mpc$^{-1}$, which also agrees well with our findings. Results from TRGB from the CCHP, EDD, and CATs and \citep[][respectively]{Freedman19, Anand22, Scolnic23} sit under 1$\sigma$ from our measurement. 

We can also test the hypothesis that the local measurement of $H_0$ is greater than the early-Universe value. The null hypothesis states that our measurement does not exceed \textit{Planck's}. We then determine the probability of rejecting that null hypothesis,
\begin{equation}
    P(H_0 \leq H_{0, Planck}) = \int_{-\infty}^{+\infty} \mathcal{P}(H)\left[ \int_{-\infty}^{H} \mathcal{P}(H_0') dH_0'\right] dH,
\end{equation}
where $\mathcal{P}(H)$ and $\mathcal{P}(H_0')$ are the posterior probabilities of the \textit{Planck} measurement (assumed to be Gaussian with mean and standard deviation from $H_{0, Planck} = 67.4 \pm 0.5$ km s$^{-1}$ Mpc$^{-1}$) and our measurement, respectively. As a result, we find that there is a $<5\%$ chance that our value is lower than \textit{Planck's} \cite{Planck18VI}. On the other hand, repeating this analysis but with a comparison to SH0ES instead, we find that there virtually no preference for a higher or lower value. Overall, thus indicates that our result reinforces the current tension ( i.e., that the local value of $H_0$ exceeds the CMB-based value) with $95\%$ confidence. Greater precision will be necessary for more definitive results in this fast evolving field. 

\section{Summary and Conclusions}\label{sec:conclusions}

We use a combination of recent and archival HST WFC3/IR \textit{F110W} and \textit{F160W} observations to discover and characterize Miras in the giant spiral galaxy M101, host to SN 2011fe. The combination of multiple observing campaigns --- some originally obtained to study the late-time light curve of SN 2011fe --- results in a baseline of up to $\sim 2900$ days for regions of the supernova field with the maximum observational overlap. From an initial sample of $\sim 3000$ candidate variable stars, we create a list of 288 Oxygen-rich Mira candidates with derived periods ranging from $200-500$ days.  We then use the dependence of zeropoint as a function of period to help determine upper (potential faint bias due to C-rich Mira contamination) and lower (potential bright bias due to incompleteness) period bounds of 240 to 400 days for the final fit. The 211 Miras with periods within this range are then used to fit a Period-Luminosity Relation to this galaxy. 

Combined with the absolute calibration from the Large Magellanic Cloud and NGC 4258 samples obtained in \cite{Huang18}, these Miras are then used to derive an independent distance measurement to M101 (the first with Mira variables) of $\mu_{\rm M101} = \ $ \mugal \ mag which is in 1$\sigma$ agreement with the weighted average of recent literature Cepheid and Tip of the Red Giant Branch distances to this galaxy. 

We also use the Mira measurement to calibrate the luminosity of its SN~Ia, SN 2011fe, and find $M^{0}_{B,\rm M101} = -19.294 \pm 0.132 $ mag. This is the second SN~Ia host galaxy, after NGC 1559 \citep{Huang_2020}, with a Mira distance. When taking the weighted average of both SN~Ia peak luminosities, we find $M^0_B = -19.268 \pm 0.088$ mag. With the intercept of the Hubble diagram, $a_B = 0.714158\pm 0.0085$ \citep{Riess22}, determined from 277 supernovae from the Pantheon+ sample \citep{Scolnic_2022, Brout_2022a}, we find H$_0 =$ \hnaught \ km$^{-1}$s$^{-1}$ Mpc, a 4.1\% measurement of $H_0$ using only geometry, Miras, and SNe Ia. This is an approximately $\sim 1/\sqrt{2}$ reduction in uncertainty from the previous result of \cite{Huang_2020} and is consistent with the measurement being dominated by the statistical uncertainty in the peak SN~Ia magnitude. 

The $H_0$ value obtained here, H$_0=72.37 \pm 2.97 { \ \rm km \ s}^{-1} {\ \rm Mpc}^{-1}$, agrees well with the baseline SH0ES measurement of $73.04 \pm 1.04$ km s$^{-1}$ Mpc$^{-1}$ as well as the NGC 4258 + LMC only value of $73.35 \pm 1.17$ km s$^{-1}$ Mpc$^{-1}$ \citep{Riess22}.  Our measurement also agrees to within 1$\sigma$ with the value obtained by CCHP using Tip of the Red Giant Branch, $H_0 = 69.8 \pm 1.9$ km s$^{-1}$ Mpc$^{-1}$ and thus does not definitively address the smaller local disagreement between the Cepheid-based SH0ES and the TRGB-based CCHP measurements. However, the result derived here does corroborate previous findings that the local value is higher than the early-Universe \textit{Planck} result at a 95\% confidence level.


\begin{acknowledgments}
CDH thanks the anonymous referee for their helpful comments and suggestions and Andrea Sacchi and Catherine Zucker for their helpful discussions and is deeply grateful to Odin Snowfoot and Nikhil Anand for their unwavering support during the writing of this paper. PAW and JWM acknowledge research support from the National Research Foundation. Support for \textit{HST} programs AR-16132 and GO-16744 was provided by NASA through a grant from the Space Telescope Science Institute, which is operated by the Association of Universities for Research in Astronomy, Inc., under NASA contract NAS5-26555. This research has made use of Astropy, a community-developed core Python package of Astronomy \citep{Astropy_2018} and data obtained from the Mikulski Archive for Space Telescopes (MAST) at the Space Telescope Science Institute. 
\end{acknowledgments}

%

\vspace{5mm}
\facilities{HST(WFC3/IR)}


\software{Astropy \citep{Astropy_2013, Astropy_2018}}, Matplotlib \citep{Hunter_2007}, NumPy \citep{Harris_2020}, DAOPHOT \cite{Stetson87, Stetson94}



\bibliography{main_new}{}
\bibliographystyle{aasjournal}



\end{document}